\documentclass[journal]{IEEEtran}

\usepackage[final]{graphicx}
\usepackage{times}
\usepackage{amsfonts}
\usepackage{epsfig}
\usepackage{amssymb}
\usepackage{amsmath}
\usepackage{citesort}
\usepackage{psfrag}
\usepackage{balance}

\newcommand{\abar}{\bar{a}}
\newcommand{\du}{\mathbf{d}}
\newcommand{\lu}{\mathbf{l}}
\newcommand{\Ca}{C_\mathrm{a}}
\newcommand{\Cae}{\mathcal{C}_\mathrm{a}}
\newcommand{\Cb}{C_\mathrm{b}}
\renewcommand{\H}{\mathbb{H}}

\renewcommand{\u}{\mathbf{u}}
\newcommand{\x}{\mathbf{x}}
\newcommand{\C}{\mathcal{C}}
\newcommand{\dmin}{d_{\rm min}}
\DeclareMathOperator{\coef}{coeff}
\DeclareMathAlphabet{\mathbit}{OT1}{cmr}{bx}{it}

\newtheorem{lemma}{Lemma}
\newtheorem{corollary}{Corollary}

\newtheorem{theorem}{Theorem}

\begin{document}
\title{On the Analysis of Weighted Nonbinary Repeat Multiple-Accumulate Codes}

\author{Eirik Rosnes,~\IEEEmembership{Senior Member,~IEEE,} and Alexandre~Graell~i~Amat,~\IEEEmembership{Senior Member,~IEEE}
\thanks{The material in this paper was presented in part at the $6$th
International Symposium on Turbo Codes \& Iterative Information
 Processing, Brest, France, September 2010, and at the Information Theory and Applications (ITA) workshop, La Jolla, CA, February 2011. The work of E. Rosnes was
supported by the Research Council of Norway (NFR) under Grants
174982 and 183316.

E. Rosnes is with the Selmer Center, Department of Informatics,
University of Bergen, N-5020 Bergen, Norway (e-mail:
eirik@ii.uib.no).

A. Graell i Amat is with the Department of Signals and Systems, Communication Systems Group,  Chalmers
University of Technology, Gothenburg, Sweden (e-mail: alexandre.graell@chalmers.se).}}


\maketitle

\begin{abstract}
In this paper, we consider weighted nonbinary repeat
multiple-accumulate (WNRMA) code ensembles obtained from the serial
concatenation of a nonbinary rate-$1/n$ repeat code and the cascade
of $L\geq 1$ accumulators, where each encoder is followed by a
nonbinary random weighter. The WNRMA codes are assumed to be
iteratively decoded using the turbo principle with maximum \emph{a
posteriori} constituent decoders. We derive the exact weight
enumerator of nonbinary accumulators and subsequently give the
weight enumerators for WNRMA code ensembles. We formally prove that
the symbol-wise minimum distance of WNRMA code ensembles
asymptotically grows linearly with the block length when $L \geq 3$
and $n \geq 2$, and $L=2$ and $n \geq 3$, for all powers of primes
$q \geq 3$ considered, where $q$ is the field size. Thus, WNRMA code
ensembles are \emph{asymptotically good} for these parameters. We
also give iterative decoding thresholds, computed by an extrinsic
information transfer chart analysis,  on the $q$ary symmetric
channel to show the convergence properties. Finally, we consider the
binary image of WNRMA code ensembles and compare the asymptotic
minimum distance growth rates with those of binary repeat
multiple-accumulate code ensembles.
\end{abstract}


\section{Introduction} \label{sec:intro}

Weighted nonbinary repeat accumulate (WNRA) codes were introduced by
Yang in \cite{yan04} as the $q$ary generalization of the celebrated
binary repeat accumulate (RA) codes. The encoder consists of a rate
$R_\mathrm{rep}=1/n$ nonbinary repeat code, a weighter, a random
symbol interleaver, and an accumulator over a finite field GF$(q)$
of size $q$. WNRA codes can be decoded iteratively using the turbo
principle, and in \cite{yan04} simulation results were presented
that showed that these codes are superior to binary RA codes on the
additive white Gaussian noise (AWGN) channel when the weighter is properly
chosen. In a recent work \cite{kim09}, Kim \emph{et al.} derived an
approximate \emph{input-output weight enumerator} (IOWE) for the
nonbinary accumulator. Based on that, approximate upper bounds on
the maximum-likelihood (ML) decoding threshold of WNRA codes  with $q$ary orthogonal modulation and coherent detection
over the AWGN channel were
computed for different values of the repetition factor $n$ and the
field size $q$, showing that these codes perform close to capacity
under ML decoding for large values of $n$ and $q$.

In \cite{pfi03}, Pfister showed that the minimum distance ($\dmin$)
of binary repeat multiple-accumulate (RMA) codes, built from the
concatenation of a repeat code with two or more accumulators,
increases as the number of accumulators increase. In particular, it
was shown in \cite{pfi03} that there exists a sequence of RMA codes
with $\dmin$ converging in the limit of
infinitely many accumulators to the Gilbert-Varshamov bound (GVB).
The stronger result that the typical $\dmin$ converges to the GVB
was recently proved in \cite{fan09}. Also, in \cite{PfThs03}, it was
conjectured by Pfister that the $\dmin$ of RMA codes asymptotically
grows linearly with the block length, and that the growth rate is
given by the threshold where the asymptotic spectral shape function
becomes positive. More recently, it was shown in \cite{all07,fan09}
that RMA code ensembles with two or more accumulators are indeed
\emph{asymptotically good}, in the sense that their $\dmin$
asymptotically grows linearly with the block length. A formal proof
was given in \cite{fan09}, and a method for the calculation of a
lower bound on the growth rate coefficient was given in
\cite{all07}. 

In a recent paper \cite{ros10_istc}, the authors considered weighted
nonbinary repeat multiple-accumulate (WNRMA) code ensembles obtained
from the serial concatenation of a nonbinary repeat code and the
cascade of $L\geq 1$ accumulators, where each encoder is followed by
a nonbinary weighter, as the $q$ary generalization of binary RMA
codes \cite{pfi03,PfThs03,all07,fan09,GraRos09}. Building upon the
approximate IOWE for nonbinary accumulators \cite{kim09}, it was
shown numerically in \cite{ros10_istc} that the $\dmin$ of WNRMA
code ensembles grows linearly with the block length, and the growth
rates were estimated. However, no formal proof was provided in
\cite{ros10_istc}. In this paper, we address this issue. We derive
an \emph{exact} expression for the IOWE of a nonbinary accumulator
which allows us to derive an exact closed-form expression for the
average \emph{weight enumerator} (WE) of WNRMA code ensembles. We
then analyze the asymptotic behavior of the average WE of WNRMA code
ensembles, extending the asymptotic $\dmin$ analysis in
\cite{all07,fan09} for binary RMA code ensembles to WNRMA code
ensembles. In particular, we prove that the $\dmin$ of WNRMA code
ensembles asymptotically grows linearly with the block length when $L \geq 3$ and $n
\geq 2$, and $L=2$ and $n \geq 3$,  for
all powers of primes $q \geq 3$ considered. Hence, WNRMA code ensembles are
asymptotically good for these parameters. The obtained growth rates
are very close to the GVB for practical values of $q$. However, for large values of $q$, the growth rate coefficient decreases with $q$, and the gap to the GVB starts to increase. Furthermore,
we consider extrinsic information transfer (EXIT) charts
\cite{ten01} to analyze the convergence properties of WNRMA codes on
the $q$ary symmetric channel (QSC). Finally, we also consider the
binary image of WNRMA codes. We give an expression for the average
binary WE of nonbinary WNRMA code ensembles and analyze its
asymptotic behavior. We also compute the asymptotic $\dmin$ growth
rates of the binary image of WNRMA code ensembles and compare them
with those of binary RMA code ensembles. For given $n$, we show that
the growth rate improves with the value of $q$ for the considered
values of $q$. Also, we compute ML decoding thresholds of the binary
image of WNRMA code ensembles on the AWGN channel and show that
these codes perform very close to capacity under ML decoding.

Nonbinary codes of low rate are potentially useful in image
watermarking applications. See, for instance, \cite{bri08} where
low-rate nonbinary turbo codes were proposed for this application.

The remainder of the paper is organized as follows. In
Section~\ref{sec:WE}, we describe the encoder structure of WNRMA
codes. We also derive an exact expression for the IOWE of a
nonbinary accumulator and a closed-form expression for the average
WE of WNRMA code ensembles. In Section~\ref{sec:AsymptAnalysis}, we
analyze the asymptotic behavior of the average WE of WNRMA code
ensembles and prove that its $\dmin$ grows linearly with the block
length. Convergence properties under iterative decoding are studied
in Section~\ref{sec:exit}, where an EXIT chart analysis is
performed. In Section~\ref{sec:BinaryImage}, we consider the binary
image of WNRMA code ensembles and compare the $\dmin$ growth rates
with those of binary RMA code ensembles. We also derive ML decoding
thresholds for these ensembles. Finally,
Section~\ref{sec:conclusion} draws some conclusions.

\section{Encoder Structure and Weight Enumerators}
\label{sec:WE}

\begin{figure}[t]
\psfrag{C0}{\small $C_0$} \psfrag{C1}{\small $C_1$}
\psfrag{CL}{\small $C_L$} \psfrag{K}{\small $K$} \psfrag{w}{\small
$w$} \psfrag{qw}{\small $nw$} \psfrag{N}{\small $N$}
\psfrag{h}{\small $h$} \psfrag{h1}{\small $h_1$} \psfrag{p1}{\small
$\pi_1$} \psfrag{p2}{\small $\pi_L$} \psfrag{Wei}{\small Random Weighter (RW)}
\centerline{\includegraphics[width=3.25in]{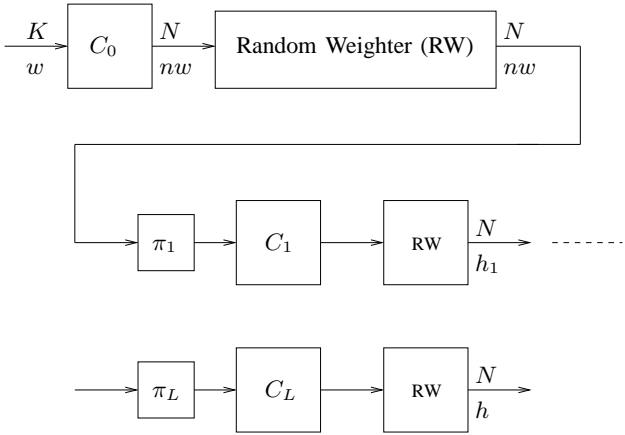}}
    \caption{Encoder structure for WNRMA codes.}
\vspace{-2mm}
    \label{fig:encoderWNRMA}
\end{figure}

The encoder structure of WNRMA codes is depicted in
Fig.~\ref{fig:encoderWNRMA}. It is the serial concatenation of a
rate $R_\mathrm{rep}=1/n$ repetition code $C_\mathrm{rep}$, with the
cascade of $L \geq 1$ identical rate-$1$, memory-one, $q$ary
accumulators ${C}_l$, $l = 1,\dots,L$, with generator polynomials
$g(D)=1/(1+D)$ over a finite field GF$(q)$, through random
interleavers $\pi_1,\dots,\pi_{L}$. Each encoder is followed by a
nonbinary weighter, which multiplies each symbol at its input by a
nonzero $q$ary symbol. For analysis purposes we consider random
weighters (RWs). We denote by $C_0$ the $(nK,K)$ outer block code
obtained by concatenating together $K$ successive codewords of
$C_\mathrm{rep}$. The overall nominal code rate (avoiding
termination) is denoted by $R=K/N=1/n$, where $N=nK$ is the output
block length. In more detail, a length-$K$ information sequence
$\mathbf{u}_0=(u_{0,1},\dots,u_{0,K})$ of $q$ary symbols
$u_{0,i}\in\{0,1,\ldots,q-1\}$ is encoded by a $q$ary repeat code.
The output of the repeat code
$\mathbf{x}_0=(x_{0,1},\dots,x_{0,nK})$ is fed to a nonbinary
weighter which multiplies each symbol $x_{0,i}$ by a nonzero $q$ary
symbol. In \cite{yan04}, it was shown that a careful choice of the
weighter can significantly improve performance. The resulting
sequence is encoded by a chain of $L$ nonbinary accumulators,
preceded by interleavers $\pi_1,\ldots,\pi_L$. Furthermore, each
accumulator is followed by a nonbinary RW.

\subsection{Average WEs for WNRMA Code Ensembles}

Let $\bar{a}^{\mathcal{C}}_{w,h}$ be the ensemble-average nonbinary
IOWE of the code ensemble $\mathcal{C}$ with input and output block
length $K$ and $N$, respectively, denoting the average number of
codewords of input Hamming weight $w$ and output Hamming weight $h$
over $\mathcal{C}$. Here, by Hamming weight, we mean the number of
nonzero symbols in a codeword. For convenience, we may simply speak
of weight. Also, denote by $\bar{a}_h^{\mathcal{C}}=\sum_{w=0}^K
\bar{a}_{w,h}^{\mathcal{C}}$ the ensemble-average nonbinary WE of
the code ensemble $\mathcal{C}$, giving the average number of
codewords of weight $h$ over $\mathcal{C}$. Throughout the paper we
will simply speak of IOWE and WE, avoiding the term nonbinary,
when the fact that they refer to nonbinary distributions is clear
from the context.

Benedetto \emph{et al.} introduced in \cite{BDMP98} the concept of
\textit{uniform interleaver} to obtain average WEs for concatenated
code ensembles from the WEs of the constituent encoders. Since
we are dealing with nonbinary codes, we need to extend the approach from \cite{BDMP98}  to consider vector-WEs. In particular, consider
the ensemble of serially concatenated codes (SCCs) obtained by
connecting two nonbinary encoders $\Ca$ and $\Cb$ through a uniform
interleaver. The ensemble-average IOWE of the serially concatenated code ensemble can be
written as
\begin{equation}\label{eq:vectorIOWE}
\abar^{\mathrm{SCC}}_{w,h}=\sum_l\sum_{\lu:\sum_{i=1}^{q-1}{l_i=l}}\frac{a^{\Ca}_{w,\lu}a^{\Cb}_{\lu,h}}{{N\choose
l_1,l_2,\ldots,\l_{q-1}}}
\end{equation}
where
\begin{displaymath}
{{N\choose l_1,l_2,\ldots,\l_{q-1}}} = \frac{N!}{l_1! \cdots l_{q-1}! (N-\sum_{i=1}^{q-1} l_i)!},
\end{displaymath}
$\lu=(l_1,l_2,\ldots,l_{q-1})$ is the weight vector with
entries $l_i$ giving the number of symbols $i$ in a codeword $\x$,
and $a^{\Ca}_{w,\lu}$ is the vector-IOWE of encoder $\Ca$, giving
the number of codewords of input weight $w$ at the input of $\Ca$
and output vector-weight $\lu$ at the output of $\Ca$, i.e., the
codeword has $l_1$ $1$'s, $l_2$ $2$'s, and so on. Likewise,
$a^{\Cb}_{\lu,h}$ is the vector-IOWE of encoder $\Cb$ giving the
number of codewords of input vector-weight $\lu$ and output weight
$h$. In general, it is very difficult to compute the vector-IOWE of
an encoder in closed-form. However, if encoder $\Ca$ is followed by
a nonbinary RW, the following theorem holds.
\begin{theorem}\label{the:vIOWE_IOWE}
Let $\C$ be the ensemble of codes over GF($q$) obtained by the
serial concatenation of two nonbinary encoders $\Ca$ and $\Cb$
through a uniform interleaver. Furthermore, encoder $\Ca$ is
followed by a nonbinary RW. Also, denote by $a^{\Ca}_{w,h}$ and
$a^{\Cb}_{w,h}$ the IOWE of encoder $\Ca$ and encoder $\Cb$,
respectively. The ensemble-average IOWE of the ensemble $\C$ can be
written as
\begin{equation}\label{eq:IOWE_SCC}
\abar^{\C}_{w,h}=\sum_l\frac{a^{\Ca}_{w,l}a^{\Cb}_{l,h}}{{N\choose
l}(q-1)^l} .
\end{equation}
\end{theorem}
\begin{IEEEproof}
Denote by $\mathcal{C}_\mathrm{a}'$ the ensemble obtained by joining
together encoder $\Ca$ and the RW. Using the concept of uniform
interleaver, the ensemble-average IOWE of the ensemble $\C$ can be
written as (see (\ref{eq:vectorIOWE}))
\begin{equation}\label{eq:vectorIOWEth}
\abar^{\C}_{w,h}=\sum_l\sum_{\lu:\sum_{i=1}^{q-1}{l_i=l}}\frac{\abar^{\Cae'}_{w,\lu}a^{\Cb}_{\lu,h}}{{N\choose
l_1,l_2,\ldots,\l_{q-1}}}
\end{equation}
where $\abar^{\Cae'}_{w,\mathbf{l}}$ is the average vector-IOWE of
the ensemble of weighted codes $\Ca$, weighted through the RW.

The average vector-IOWE of the ensemble $\Cae'$ can be written as a
function of the vector-IOWEs of encoder $\Ca$ and of the RW as
\begin{equation}\label{eq:vectorIOWECaprime}
\abar^{\Cae'}_{w,\lu}=\sum_{\du}a^{\Ca}_{w,\du}a^{\mathrm{RW}}_{\du,\lu}.
\end{equation}

Notice that the RW (over GF($q$)) is such that the weight is
preserved, i.e., $a^{\mathrm{RW}}_{\du,\lu}$ is nonzero if and only if $\sum_{i=1}^{q-1}d_i=\sum_{i=1}^{q-1}l_i$.
Therefore, we can rewrite (\ref{eq:vectorIOWECaprime}) as
\begin{equation} \notag 
\abar^{\Cae'}_{w,\lu}=\sum_{\du:\sum_{i=1}^{q-1}d_i=\sum_{i=1}^{q-1}l_i}a^{\Ca}_{w,\du}a^{\mathrm{RW}}_{\du,\lu}.
\end{equation}
Notice also that the following property holds for a nonbinary random
(uniform) weighter:
\begin{equation} \notag 
a^{\mathrm{RW}}_{\du,\lu}=a^{\mathrm{RW}}_{\du',\lu}\;\;\forall\,
\du,\du'~\mathrm{such}~\mathrm{that}~\sum_{i=1}^{q-1}d_i=\sum_{i=1}^{q-1}d'_i.
\end{equation}
In other words, the vector-IOWE of the RW depends only on the weight
$l=\sum_{i=1}^{q-1}l_i$, and we can write
\begin{equation}\label{eq:eqvRW_RW}
a^{\mathrm{RW}}_{\du,\lu}=a^{\mathrm{RW}}_{l,\lu}\;\;\forall\,
\du~\mathrm{such}~\mathrm{that}~\sum_{i=1}^{q-1}d_i=l.
\end{equation}
It is easy to verify that the vector-IOWE $a^{\mathrm{RW}}_{l,\lu}$
is given by
\begin{equation}\label{eq:vectorIOWE_RW2}
a^{\mathrm{RW}}_{l,\lu}=\frac{{l\choose
l_1,l_2,\ldots,l_{q-1}}}{(q-1)^l}.
\end{equation}
\begin{figure*}[!t]
\normalsize 
\setcounter{equation}{9}
\begin{equation}\label{eq:CWE_WNRMA}
\begin{split}
\bar{a}^{\mathcal{C}_{\mathrm{WNRMA}}}_{w,h_1,\dots,h_{L-1},h} &=
\frac{{K \choose w} (q-1)^{w} \prod_{l=1}^L
\sum_{k_l=\max(1,h_{l-1}-h_l)}^{\lfloor h_{l-1}/2 \rfloor} {N-h_l
\choose k_l} {h_l-1 \choose k_l-1} {h_{l}-k_l \choose h_{l-1}-2k_l}
\left( q-1 \right)^{k_l} \left( q-2 \right)^{h_{l-1}-2k_l}}{
\prod_{l=1}^L {N \choose h_{l-1}}(q-1)^{h_{l-1}}} \\
&= \sum_{k_1=\max(1,h_0-h_1)}^{\lfloor h_0/2 \rfloor} \sum_{k_2=\max(1,h_1-h_2)}^{\lfloor h_1/2 \rfloor} \cdots \sum_{k_L=\max(1,h_{L-1}-h_L)}^{\lfloor h_{L-1}/2 \rfloor}  \bar{a}^{\mathcal{C}_{\rm WNRMA}}_{w,h_1,\dots,h_{L-1},k_1,\dots,k_L,h}
\end{split}
\end{equation}
\setcounter{equation}{6}
\vspace*{-2mm}
\end{figure*}%
Finally, using (\ref{eq:vectorIOWE_RW2}), (\ref{eq:eqvRW_RW}),
(\ref{eq:vectorIOWECaprime}), and the fact that
\begin{equation} \notag
\sum_{\du:\sum_{i=1}^{q-1}d_i=l}a^{\Ca}_{w,\du}=a^{\Ca}_{w,l}\; \text{ and }\;
\sum_{\lu:\sum_{i=1}^{q-1}l_i=l}a^{\Cb}_{\lu,h}=a^{\Cb}_{l,h}
\end{equation}
in (\ref{eq:vectorIOWEth}), after some simple manipulations, we obtain
(\ref{eq:IOWE_SCC}), which completes the proof.
\end{IEEEproof}

From Theorem~\ref{the:vIOWE_IOWE} it follows that the
ensemble-average IOWE of WNRMA code ensembles can be computed, when
each constituent encoder is followed by a nonbinary RW, from the
IOWEs of the component encoders, which are easier to compute in
closed-form than the vector-IOWEs. Using
Theorem~\ref{the:vIOWE_IOWE} and the concept of uniform interleaver,
the ensemble-average IOWE of a WNRMA code ensemble $\mathcal{C}_{\rm
WNRMA}$ can be written as
\begin{equation} \label{eq:IOWE_WNRMA1}
\begin{split}
\bar{a}_{w,h}^{\mathcal{C}_{\rm WNRMA}} &= \sum_{h_1=0}^N \cdots \sum_{h_{L-1}=0}^N \frac{a_{w,nw}^{C_0} a_{nw,h_1}^{C_1}}{ {N \choose nw} (q-1)^{nw}} \\
&\;\;\;\; \times \left[ \prod_{l=2}^{L-1} \frac{a_{h_{l-1},h_l}^{C_l}}{ {N \choose h_{l-1}}(q-1)^{h_{l-1}}} \right]  \frac{a_{h_{L-1},h}^{C_L}}{{N \choose h_{L-1}}(q-1)^{h_{L-1}}} \\
&= \sum_{h_1=0}^N \cdots \sum_{h_{L-1}=0}^N \bar{a}^{\mathcal{C_{\rm
WNRMA}}}_{w,h_1,\dots,h_{L-1},h}
\end{split}
\end{equation}
where
$\bar{a}^{\mathcal{C}_{\mathrm{WNRMA}}}_{w,h_1,\dots,h_{L-1},h}$ is
called the \textit{conditional weight enumerator} (CWE) of
$\mathcal{C}_{\mathrm{WNRMA}}$.

The evaluation of (\ref{eq:IOWE_WNRMA1}) requires the computation of
the IOWEs of the constituent encoders, which is addressed below.

\subsection{IOWEs for Memory-One Encoders and the Repetition Code}

An approximated expression for the IOWE of a $q$ary accumulator was
given in \cite{kim09}. In this section, we derive the exact
expression for the IOWE of a $q$ary accumulator.
\begin{theorem}
The IOWE for rate-$1$, memory-one, $q$ary convolutional encoders
over GF($q$) with generator polynomials $g(D)=1/(1+D)$ and
$g(D)=1+D$ that are terminated to the zero state at the end of the
trellis and with input and output block length $N$ can be given in
closed form as
\begin{equation}\label{eq:IOWEacc1}
\begin{split}
a_{w,h}^{\frac{1}{1+D}}=a_{h,w}^{1+D}=\sum_{k=\max(1,w-h)}^{\left\lfloor
w/2 \right\rfloor}
&\binom{N-h}{k} \binom{h-1}{k-1} \binom{h-k}{w-2k}\\
&\times\left(q-1 \right)^k \left(q-2 \right)^{w-2k}
\end{split}
\end{equation}
for positive input weights $w$, where $k$ is the number of error
events. Also, $a_{0,0}^{\frac{1}{1+D}}= a_{0,0}^{1+D}=1$.
\end{theorem}
\begin{IEEEproof}
Consider a nonbinary encoder $C$ with input and output length $N$.
Denote by an error event a path through the trellis which diverges
from the all-zero state at depth $t_\mathrm{i}$ and merges again with
the all-zero state at depth $t_\mathrm{f}$, where
$t_\mathrm{f}>t_\mathrm{i}$. A nonzero codeword of input weight $w$
and output weight $h$ corresponds to the concatenation of $k$ error
events with a total input weight $w$ and a total output weight $h$.
Partition all the error events into equivalence classes based on
their length (or, equivalently, based on their accumulated output
weight). In particular, all error events within a specific class are
required to have the same length. By considering only classes of
events (i.e., we do not distinguish between error events within the
same class), $k$ error events with an accumulated output weight $h$
can be concatenated (without overlapping) in
\begin{displaymath}
\binom{N-h}{k} \binom{h-1}{k-1}
\end{displaymath}
different ways.

The next step is to consider all the error events within the same
class. First, take a look at the structure of the error events.
Notice that the first transition of an error event (the one
diverging from the all-zero state) has always input weight one and
output weight one, while the last transition of an error event (the
one merging with the all-zero state) has always input weight one and
output weight zero. Thus, the total input weight accumulated at the
boundaries (first and last transition) of the error events is $2k$,
while the total output weight accumulated at the boundaries of the
error events is $k$. Now, each error event has $q-1$ possibilities
for the first transition (since the edge from the all-zero state to
the all-zero state is not allowed), therefore, overall, we have
$(q-1)^k$ possibilities. On the other hand, there is only a single
possibility for the last transition (the one merging with the all-zero
state). Finally, we must distribute the remaining input weight,
$w-2k$, in the $h-k$ remaining transitions (i.e., excluding the
boundaries) of the error events, resulting in
\begin{displaymath}
\binom{h-k}{w-2k}
\end{displaymath}
possible distributions for the remaining input weight. Furthermore,
for each of the nonzero input weight transitions, we have $q-2$
additional possibilities (we must exclude the edge of input weight zero
and the edges merging with the all-zero state), resulting in
$(q-2)^{w-2k}$ possibilities in total. Thus, overall there are
\begin{displaymath}
\binom{N-h}{k} \binom{h-1}{k-1}(q-1)^k\binom{h-k}{w-2k}(q-2)^{w-2k}
\end{displaymath}
codewords of input weight $w$ and output weight $h$ resulting from
the concatenation of $k$ error events. The result for the encoder
$g(D) = 1/(1 + D)$ in (\ref{eq:IOWEacc1}) follows by summing over
all possible values of $k$. The IOWE for the feedforward encoder
with generator polynomial $g(D) = 1 + D$ is obtained in a similar
manner.
\end{IEEEproof}

Notice that the formula in (\ref{eq:IOWEacc1}) generalizes the
closed-form expression for the IOWE for rate-$1$, memory-one, binary
convolutional encoders from \cite{div98} to the $q$ary case.
\begin{theorem}
The IOWE for the $(nK,K)$ $q$ary repetition code $C_0$ with input
block length $K$ can be given in closed form as
\begin{equation} \label{eq:rep}
a^{C_0}_{w,nw} = {K \choose w} (q-1)^w.
\end{equation}
\end{theorem}

\begin{IEEEproof}
The number of binary vectors of length $K$ and weight $w$ is ${K
\choose w}$, and the result follows by multiplying this number by
$w$ times the number of nonzero elements from GF($q$).
\end{IEEEproof}

Using (\ref{eq:IOWEacc1}) and (\ref{eq:rep}) in
(\ref{eq:IOWE_WNRMA1}), we get the expression (\ref{eq:CWE_WNRMA})
at the top of the page for the CWE (with $w>0$) of WNRMA code
ensembles, where for conciseness $h_0=nw$ and $h_L=h$.

\section{Asymptotic Analysis of the Minimum Distance}
\label{sec:AsymptAnalysis}

With regard to (\ref{eq:CWE_WNRMA}) at the top of the page, without
loss of generality we can write \setcounter{equation}{10}
\begin{align}
   w &= \alpha N^a, &h_i &= \beta_i N^{b_i}, i=1,\ldots,L-1,\notag\\
   h &= \rho N^c,   &k_i &= \gamma_i N^{d_i}, i=1,\ldots,L \notag
\end{align}
where $0 \leq  a \leq b_1 \leq b_2 \leq \cdots \leq b_{L-1} \leq c
\leq 1$, $0 \leq d_1 \leq a \leq 1$, and $0 \leq d_i \leq b_{i-1}
\leq 1$, $i=2,\ldots,L$. These inequalities can be derived from the
binomial coefficients in the expression in (\ref{eq:CWE_WNRMA})
combined with the fact that for a binomial coefficient
$\binom{n}{k}$, $n \geq k \geq 0$. Also,
$\alpha,\beta_1,\ldots,\beta_{L-1},\gamma_1,\ldots,\gamma_L$, and
$\rho$ are positive constants. We must consider two cases: 1) at
least one of the quantities $w,h_1,\ldots,h_{L-1},k_1,\hdots,k_L$,
or $h$ is of order $o(N)$, and 2) all quantities
$w,h_1,\ldots,h_{L-1},k_1,\hdots,k_L$, and $h$ can be expressed as
fractions of the block length $N$, i.e.,
$a=b_1=\cdots=b_{L-1}=d_1=\cdots=d_L=c=1$. The following lemma
addresses the first case for weighted nonbinary
repeat double-accumulate (WNRAA) code ensembles.

\begin{lemma} \label{lem:lemma1}
In the ensemble of WNRAA codes with block length $N$ and $n \geq 3$,
in the case where at least one of the quantities $w$, $h_1$, $k_1$,
$k_2$, or $h$ is of order $o(N)$, $N^5
\abar^{\mathcal{C}_{\mathrm{WNRAA}}}_{w,h_1,k_1,k_2,h}
\longrightarrow 0$ as $N \longrightarrow \infty$ for all positive values of $h$.
\end{lemma}
\begin{IEEEproof}
The expression in (\ref{eq:CWE_WNRMA}) is very similar to the expression for the
conditional support size enumerating function of RMA code ensembles
 \cite[Eq.\ (10)]{GraRos09}. In particular, the
binomial coefficients in (\ref{eq:CWE_WNRMA}) are identical to those
of \cite[Eq.\ (10)]{GraRos09}. The only difference is that
(\ref{eq:CWE_WNRMA}) has some extra terms in the form of powers of
$q-1$ and $q-2$. 
Therefore, the proof of
\cite[Lemma 3]{GraRos09} applies, with some modifications, also
here.
\end{IEEEproof}

Lemma~\ref{lem:lemma1} can be generalized to the case of WNRMA code
ensembles with $L \geq 3$. The proof is omitted for brevity. As a
consequence of Lemma~\ref{lem:lemma1}, we can assume that $w$,
$h_1,\ldots,h_{L-1},k_1,\ldots,k_L$, and $h$ are all linear in the
block length: The average number of codewords of weight at most
$\hbar$, for some $\hbar$, of WNRMA code ensembles is upper-bounded
by
\begin{displaymath}
N^{2L+1} \max_{w,h_1,\dots,h_{L-1},k_1,\dots,k_L,h \leq \hbar}\abar^{\mathcal{C}_{\mathrm{WNRMA}}}_{w,h_1,\dots,h_{L-1},k_1,\dots,k_L,h}
\end{displaymath}
which from Lemma~\ref{lem:lemma1} tends to zero as $N$ tends to
infinity if at least one of the quantities is of order $o(N)$. Thus,
the average number of codewords of sublinear weight of at most
$\hbar$ tends to zero as $N$ tends to infinity.

We now address the second case by analyzing the asymptotic spectral shape
function. The asymptotic spectral shape function is defined as
\cite{Gal63}
\begin{equation} \notag 
        r(\rho) = \limsup_{N\longrightarrow\infty}\frac{1}{N}\ln\bar{a}^\mathcal{C}_{\left \lfloor\rho N \right \rfloor}
\end{equation}
where $\sup(\cdot)$ denotes the supremum of its argument, $\rho=
\frac{h}{N}$ is the normalized output weight, and $N$ is the code
block length. If there exists some abscissa $\rho_0 > 0$ such that
$\sup_{\rho\leq\rho^*}r(\rho)<0\quad\forall\rho^*<\rho_0$, and
$r(\rho)
> 0$ for some $\rho > \rho_0$, then it can be shown that, with high
probability, the $\dmin$ of most codes in the ensemble grows
linearly with the block length $N$, with growth rate coefficient of
at least $\rho_0$. On the other hand, if $r(\rho)$ is strictly zero
in the range $(0,\rho_0)$, it cannot be proved directly whether
the $\dmin$ grows linearly with the block length or not. In
\cite{fan09}, it was shown that the asymptotic spectral shape function of RMA
codes exhibits this behavior, i.e., it is zero in the range
$(0,\rho_0)$ and positive for some $\rho>\rho_0$. By combining the
asymptotic spectral shapes with the use of bounding techniques, it
was proved in \cite[Theorem 6]{fan09} that the $\dmin$ of RMA code
ensembles indeed grows linearly with the block length with growth
rate coefficient of at least $\rho_0$.

We remark that in the rest of the paper, with a slight abuse of
language, we sometimes refer to $\rho_0$ as the exact value of the
asymptotic growth rate coefficient. However, strictly speaking,
$\rho_0$ is only a lower bound on it.

Now, by using Stirling's approximation for the binomial coefficient
${n \choose k} \sim \mathrm{e}^{n\mathbb{H}(k/n)}$ for $n \to \infty$ and $k/n$ constant, where $\mathbb{H}(\cdot)$ is the binary
entropy function with natural logarithms, and the fact that $w$,
$h_1,\dots,h_{L-1},k_1,\dots,k_{L}$, and $h$ can all be assumed to be of the same order as $N$ (due to Lemma~\ref{lem:lemma1}, generalized to the general case), $\bar{a}^{\mathcal{C}_\mathrm{WNRMA}}_{w,h_1,
\ldots,h_{L-1},h}$  can be written as
\begin{equation} \notag
\begin{split}
&\bar{a}^{\mathcal{C}_\mathrm{WNRMA}}_{w,h_1, \ldots,h_{L-1},h} = \\ 
   &\sum_{k_1,\ldots,k_L}\exp \left\{ f(\alpha, \beta_1, \ldots,\beta_{L-1}, \gamma_1, \ldots,\gamma_{L}, \rho)\,N+o(N)\right\} 
\end{split}
\end{equation}
when $N\longrightarrow \infty$, where $\alpha=\frac{w}{K}$ is the
normalized input weight, $\beta_l=\frac{h_l}{N}$ is the normalized
output weight of code $C_l$, $\gamma_l=\frac{k_l}{N}$, and the
function $f(\cdot)$ is given by

\begin{equation}  \label{eq:objWNRMA}
\begin{split}
&f(\beta_0, \beta_1, \ldots,\beta_{L-1}, \gamma_1,
\ldots,\gamma_{L},
\rho) \\
&~~~~=\frac{\H \left( \beta_0 \right)}{n}  -\sum_{l=1}^L \H \left( \beta_{l-1} \right) + \sum_{l=1}^L (1-\beta_l) \H \left( \frac{\gamma_l}{1-\beta_l} \right)\\
&~~~~\;\;\;\;+\sum_{l=1}^L \beta_l \H \left( \frac{\gamma_l}{\beta_l} \right)
+\sum_{l=1}^L (\beta_{l}-\gamma_l) \H \left( \frac{\beta_{l-1}-2\gamma_l}{\beta_{l}-\gamma_l} \right) \\
&~~~~\;\;\;\;+ \ln(q-1)\sum_{l=1}^L (\gamma_l-\beta_{l-1}) \\
&~~~~\;\;\;\;+\ln(q-2) \sum_{l=1}^L(\beta_{l-1}-2\gamma_l) +
\frac{\beta_0 \ln(q-1)}{n}
\end{split}
\end{equation}
where for conciseness we defined $\beta_0=\alpha$ and
$\beta_L=\rho$.
Finally, the asymptotic spectral shape function for WNRMA code ensembles can be
written as
\begin{equation}\label{eq:spshape2}
\begin{split}
&        r^{\mathcal{C}_{\mathrm{WNRMA}}}(\rho) \\
&\;\;\;\;=  \sup_{\substack{0 \leq \beta_{l-1} \leq 1\\
\max(0,\beta_{l-1}-\beta_l) \leq \gamma_l \leq \\
\min(\beta_l,1-\beta_l,\beta_{l-1}/2)\\l=1,\dots,L}}  f(\beta_0, \beta_1, \ldots,
\beta_{L-1},\gamma_1, \ldots, \gamma_{L},\rho).
\end{split}
\end{equation}



Note that the objective function in (\ref{eq:spshape2}), defined in
(\ref{eq:objWNRMA}), can be rewritten into \cite[Eq.\
(6)]{ros10_istc}, since
\begin{displaymath}
\begin{split}
&\sum_{l=1}^L \beta_l \H \left( \frac{\gamma_l}{\beta_l} \right) +\sum_{l=1}^L (\beta_{l}-\gamma_l) \H \left( \frac{\beta_{l-1}-2\gamma_l}{\beta_{l}-\gamma_l} \right)\\
&= \sum_{l=1}^L \beta_l \H \left( \frac{\beta_{l-1}-\gamma_l}{\beta_l} \right) +\sum_{l=1}^L (\beta_{l-1}-\gamma_l) \H \left( \frac{\gamma_l}{\beta_{l-1}-\gamma_l} \right).
\end{split}
\end{displaymath}
Thus, the approximate asymptotic spectral shape function given in
\cite[Eq.\ (7)]{ros10_istc} is indeed exact. Therefore, the growth
rate coefficients computed in this section coincide with those in
\cite{ros10_istc}. However, for finite block lengths, the IOWE of a
nonbinary accumulator as given by Theorem~1 in \cite{ros10_istc}
using the approximation for $p(k)$ given in \cite[Eq.\
(3)]{ros10_istc} (which is taken from \cite{kim09}) is not exact.

From (\ref{eq:objWNRMA}) and (\ref{eq:spshape2}) it can easily be
verified that the asymptotic spectral shape function of WNRMA code
ensembles satisfies the recursive relation
\begin{displaymath}
r^{\mathcal{C}_{{\rm WNRMA}(l)}}(\rho) = \sup_{0 \leq u \leq 1}
\left[ r^{\mathcal{C}_{{\rm WNRMA}(l-1)}}(u)+ \psi(u,\rho) \right]
\end{displaymath}
where $r^{\mathcal{C}_{{\rm WNRMA}(l)}}$, $l>0$, is the asymptotic
spectral shape function with $l$ accumulators, $r^{\mathcal{C}_{\rm
WNRMA(0)}}(\rho)=\frac{1}{n}(H(\rho)+\rho\ln(q-1))$ is the
asymptotic
 spectral shape function of a repeat code, and
\begin{equation} \label{eq:f}
\begin{split}
\psi(u,\rho) &= \sup_{\substack{ \max(0,u-\rho) \leq  \gamma \leq \\
\min(\rho,1-\rho,u/2)}} \left[  -\mathbb{H}(u) + \rho
\mathbb{H} \left( \frac{\gamma}{\rho} \right) \right. \\
&~~~~\;\;\;\;+(1-\rho)\mathbb{H} \left( \frac{\gamma}{1-\rho} \right) +
\left.(\rho-\gamma) \mathbb{H} \left(
\frac{u-2\gamma}{\rho-\gamma} \right) \right. \\ 
&~~~~\;\;\;\;\left. +(\gamma-u) \ln (q-1)+(u-2\gamma) \ln(q-2) \right].
\end{split}
\end{equation}

\begin{lemma} \label{prop:1}
The asymptotic spectral shape function of the
WNRMA code ensemble is nonnegative, i.e.,
\begin{displaymath}
r^{\mathcal{C}_{\mathrm{WNRMA}(l)}}(\rho)\geq 0,~\forall
\rho\in[0,1].
\end{displaymath}
\end{lemma}

\begin{IEEEproof}
We have $r^{\mathcal{C}_{\mathrm{WNRMA}(1)}}(\rho)\geq
\psi(0,\rho)+H(0)/n=0$. The general case can be proved by induction
on $l$.
\end{IEEEproof}

\begin{figure}[!t]
\centerline{\includegraphics[width=\columnwidth]{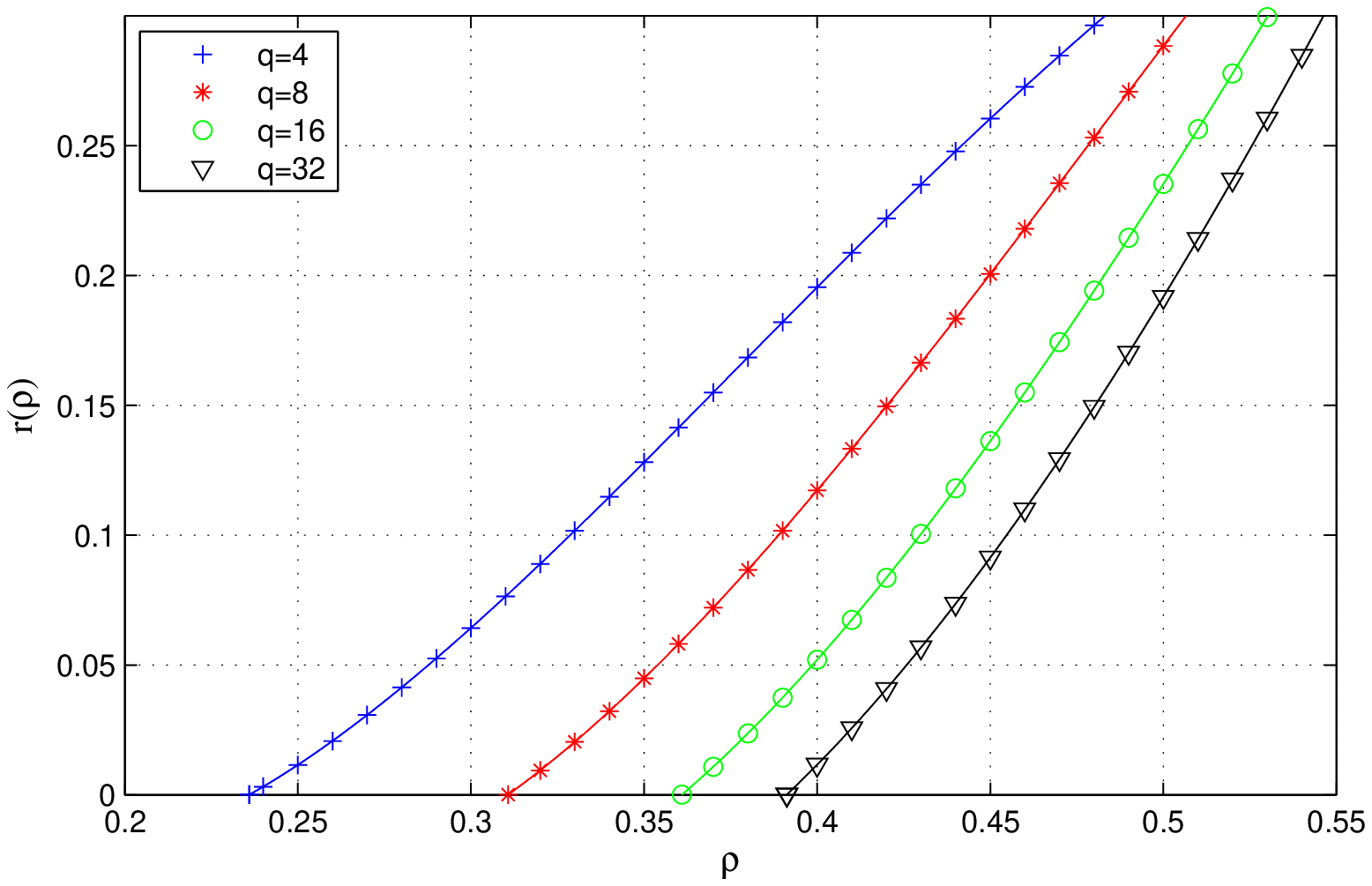}}
\vspace{-3mm} \caption{\label{fig:L2n3} {Asymptotic
spectral shape function of WNRAA codes with $n=3$.}} 
\vspace{-2mm}
\end{figure}

\begin{figure}[tbp]
\centerline{\includegraphics[width=\columnwidth]{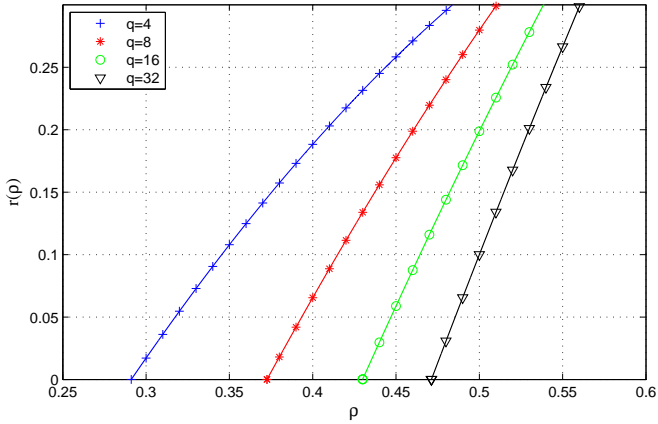}}
\vspace{-3mm} \caption{\label{fig:L3n3} {Asymptotic
spectral shape function of WNRAAA codes with $n=3$. }} \vspace{-2mm}
\end{figure}

To analyze the asymptotic $\dmin$ behavior of WNRMA code ensembles,
we must solve the optimization problem in
(\ref{eq:objWNRMA})-(\ref{eq:spshape2}). 
An efficient algorithm to
solve this problem is given in Appendix~A. The numerical evaluation
of (\ref{eq:objWNRMA})-(\ref{eq:spshape2}) is shown in
Figs.~\ref{fig:L2n3} and \ref{fig:L3n3} for WNRAA and  weighted nonbinary repeat
triple-accumulate (WNRAAA) code ensembles, respectively,  with $n=3$
and $q=4,8,16$, and $32$. The asymptotic spectral shape function
is zero in the range $(0,\rho_0)$ and positive for some
$\rho>\rho_0$. In this case, we cannot conclude directly whether the $\dmin$ asymptotically grows linearly
with the block length or not. However, we can prove the following theorem.
\begin{theorem}\label{th:rma_growth}
Define
$\rho_0=\max\{\rho^*\in[0,(q-1)/q):r^{\mathcal{C}_{\rm WNRMA}}(\rho)=0~\forall\rho\leq
\rho^*\}$. Then $\forall\rho^*>0$
\begin{displaymath}
\lim_{N\longrightarrow\infty}\mathrm{Pr}\left(d_{\rm
min}\leq(\rho_0-\rho^*)N\right)=0
\end{displaymath}
when $L \geq 3$ and $n \geq 2$, and $L=2$ and $n \geq 3$, for all powers of primes $q \geq 3$.
Thus, if $\rho_0 > 0$ and $r^{\mathcal{C}_{\rm WNRMA}}(\rho) \geq 0~\forall\rho$ (see Lemma~\ref{prop:1}),
then almost all codes in the ensemble have asymptotic
minimum distance growing linearly with $N$ with growth rate
coefficient of at least $\rho_0$.
\end{theorem}

\begin{IEEEproof}
See Appendix~B.
\end{IEEEproof}

\begin{figure}[!t]
\centerline{\includegraphics[width=\columnwidth]{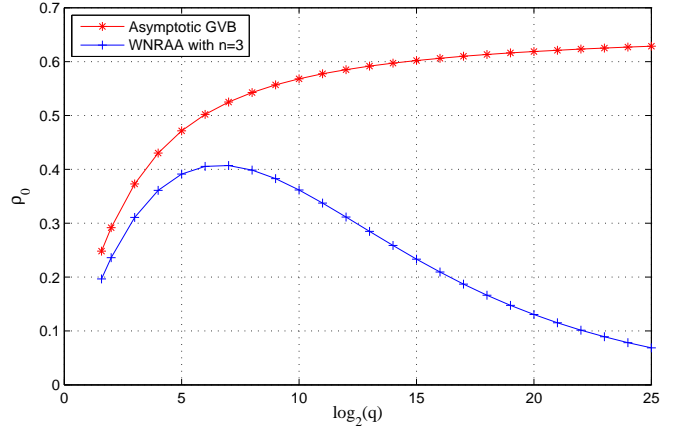}}
\caption{\label{fig:crossingversusqn3L2} {The value of $\rho_0$ versus the field size $q$ for $q=3$ and $q=2^l$, $2 \leq l \leq 25$, when $n=3$ and $L=2$.}}
\end{figure}

\begin{figure}[!t]
\centerline{\includegraphics[width=\columnwidth]{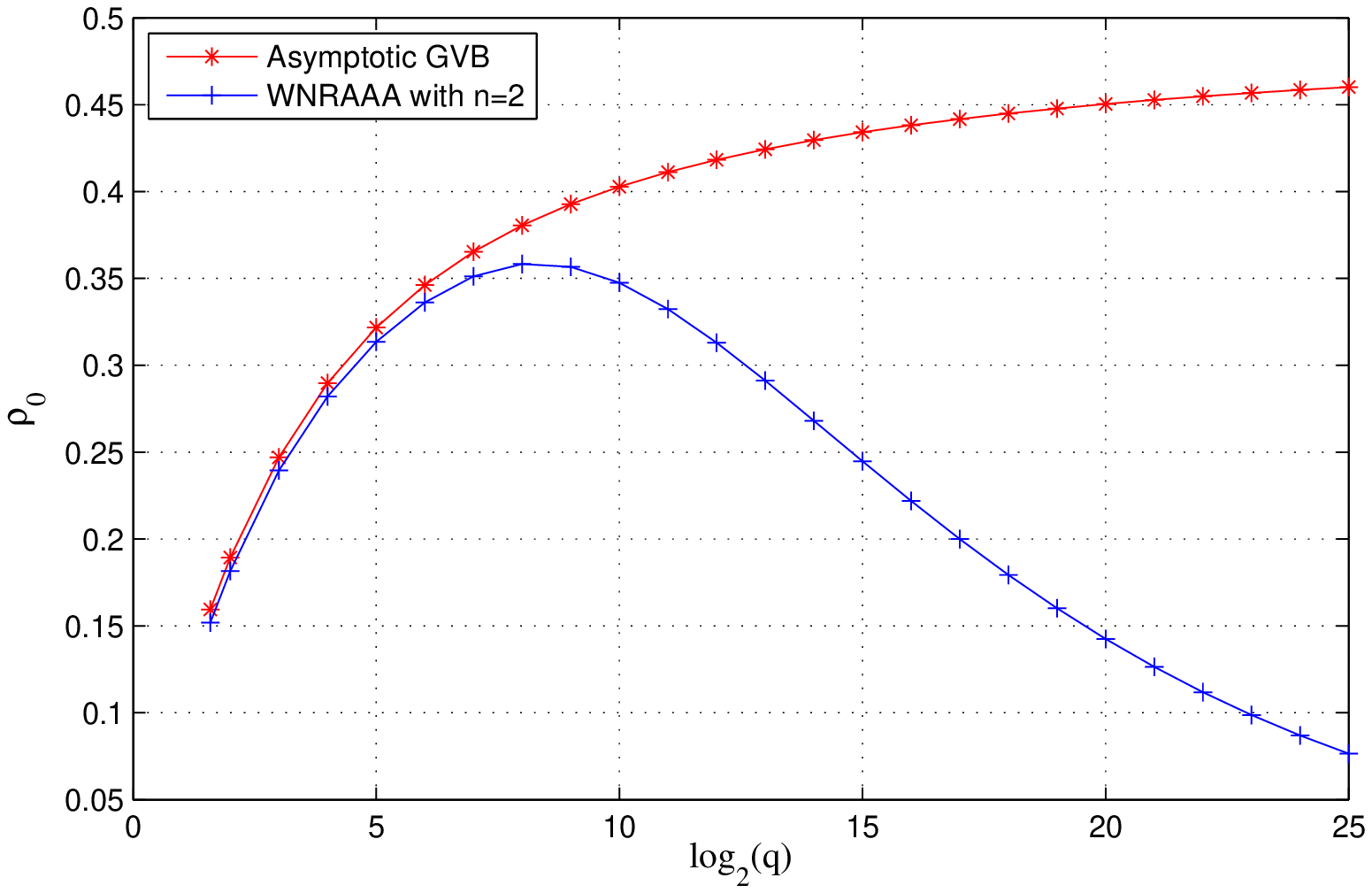}}
\caption{\label{fig:crossingversusqn2L3} {The value of $\rho_0$ versus the field size $q$ for $q=3$ and $q=2^l$, $2 \leq l \leq 25$, when $n=2$ and $L=3$.}}
\end{figure}

We can now prove the following theorem.
\begin{theorem}\label{th:rma_growth2}
The typical $\dmin$ of WNRMA code ensembles when $L \geq 3$ and $n \geq 2$, and $L=2$ and $n \geq 3$, for all powers of primes $3 \leq q \leq 2^{25}$,  grows
linearly with the block length.
\end{theorem}
\begin{IEEEproof}
The result follows from Theorem~\ref{th:rma_growth} by showing that $\rho_0$ (as defined in Theorem~\ref{th:rma_growth}) is strictly positive  when $L \geq 3$ and $n \geq 2$, and $L=2$ and $n \geq 3$,  for all powers of primes $3 \leq q \leq 2^{25}$.

 It follows directly from the definition of the objective function in (\ref{eq:objWNRMA}) that the asymptotic spectral shape function is nonincreasing in $n$, and thus $\rho_0$ is nondecreasing in $n$.

Furthermore, note that if we serially concatenate any nonbinary encoder whose
$\dmin$ grows linearly with the block length with growth
rate coefficient of at least $\rho_0$ with a nonbinary accumulate code followed by a uniform weighter through a uniform interleaver, the resulting concatenated code
ensemble will exhibit a $\dmin$ growing linearly with the
block length with growth rate coefficient of at least $\left\lceil
\rho_0/2 \right\rceil$. This follows from the fact that the output
weight $h$ of a nonbinary accumulate code is lower bounded by
$\left\lceil\frac{w}{2}\right\rceil$, where $w$ is the nonbinary input weight. This follows directly from the binomial coefficient $\binom{h-k}{w-2k}$ in (\ref{eq:IOWEacc1}), since it implies that $h-k
\geq w-2k$, from which it follows that $w \leq h+k \leq
h+\left\lfloor w/2 \right\rfloor$, which implies that $h \geq
\left\lceil\frac{w}{2}\right\rceil$.

Thus, increasing $n$ or $L$ does not change the asymptotic $\dmin$ linear growth property. The final part of the proof considers the last \emph{dimension}, i.e., what happens when $q$ increases.


By numerically solving the optimization problem in (\ref{eq:spshape2}), we find that $\rho_0=0.1966$ for $q=3$, $n=3$, and $L=2$, and $\rho_0=0.1519$ for $q=3$, $n=2$, and $L=3$. Furthermore, Figs.~\ref{fig:crossingversusqn3L2} and \ref{fig:crossingversusqn2L3} show the value of $\rho_0$ (computed numerically by solving the optimization problem in (\ref{eq:spshape2}) as function of the field size $q$ for $q=3$ and $q=2^l$, $2 \leq l \leq 25$, when $n=3$ and $L=2$, and $n=2$ and $L=3$, respectively. 
From the figures, we observe that $\rho_0$ is strictly positive for
$q \leq 2^{25}$ in both cases, i.e., for both $n=3$ and $L=2$, and
$n=2$ and $L=3$, which concludes the proof.
\end{IEEEproof}

We remark that we have limited the value of $q$ to $2^{25}$, which is much higher than any value used in practice. However, from Figs.~\ref{fig:crossingversusqn3L2} and \ref{fig:crossingversusqn2L3}, we observe that the result of Theorem~\ref{th:rma_growth2} will also hold for larger values of $q$. 

The exact values of $\rho_0$ are given in Table~\ref{tab:crossingsL2} for
several values of the repetition factor $n$ and the field size $q$
for WNRAA codes. For comparison, we
have also tabulated the asymptotic $\dmin$ growth rate coefficient
from the asymptotic GVB for nonbinary codes computed from
\begin{equation} \notag
R \geq \begin{cases} 1-\H_q(\rho_{\rm min})-\rho_{\rm min} \log_q(q-1), & \text{if $\rho_{\rm min} \leq \frac{q-1}{q}$} \\
0, & \text{otherwise} \end{cases}
\end{equation}
where $\rho_{\rm min}$ is the normalized $\dmin$, $R$ is the
asymptotic rate, and $\H_q(\cdot)$ is the binary entropy function
with base-$q$ logarithms. We observe that the gap to the GVB
decreases with increasing values of $n$ for a fixed value of $q$.
For a fixed value of $n$, the growth rate coefficient increases with
increasing values of $q$, while the gap to the GVB stays
approximately constant. However, as can be seen from Figs.~\ref{fig:crossingversusqn3L2} and \ref{fig:crossingversusqn2L3}, this behavior only holds for small values of $q$. In fact,  the asymptotic growth rate coefficient increases with the field size $q$ up to some value, and then it decreases again, after which the gap to the GVB also increases. This is also consistent with the behavior observed for nonbinary low-density parity-check codes in \cite{kas08}.
The values of $\rho_0$ for WNRAAA code ensembles are given in
Table~\ref{tab:crossingsL3} for selected values of $n$ and $q$.
The growth rate coefficients are very close
to the GVB for WNRAA code ensembles with $n=5$ and $n=10$ and for WNRAAA
code ensembles, for the considered values of $q$. For WNRAAA code ensembles with $n=5$ and $n=10$ the growth
rates coincide with the GVB, for the considered values of $q$.


\begin{table}[!t]
\setlength{\tabcolsep}{4.0pt}
\scriptsize \centering \caption{Growth rate coefficient
$\rho_0$ of WNRAA codes for different values of the repetition
factor $n$ and the field size $q$. The corresponding growth rates
from the asymptotic nonbinary GVB are given in the parentheses.}
\label{tab:crossingsL2}
\def\Hline{\noalign{\hrule height 2\arrayrulewidth}}
\vskip -3.0ex 
\begin{tabular}{lcccc}
\Hline \\ [-2.0ex]
   &   $q=4$ &  $q=8$ &  $q=16$&  $q=32$ \\
\hline
%
%
%
\\ [-2.0ex] \hline  \\ [-2.0ex]

 $n=3$  & 0.2360 (0.2917) &  0.3107 (0.3730) &  0.3609 (0.4302) & 0.3912 (0.4715)\\
 $n=5$  & 0.3820 (0.3977) &  0.4840 (0.4987) &  0.5518 (0.5664) & 0.5967 (0.6131) \\
 $n=10$ & 0.5026 (0.5048) &  0.6192 (0.6207) &  0.6930 (0.6940) & 0.7413 (0.7421) \\
\hline
\end{tabular}
\end{table}
\begin{table}[!t]
\setlength{\tabcolsep}{4.0pt}
\scriptsize \centering \caption{Growth rate coefficient
$\rho_0$ of WNRAAA codes for different values of the repetition
factor $n$ and the field size $q$. The corresponding growth rates
from the asymptotic nonbinary GVB are given in the parentheses.}
\label{tab:crossingsL3}
\def\Hline{\noalign{\hrule height 2\arrayrulewidth}}
\vskip -3.0ex 
\begin{tabular}{lcccc}
\Hline \\ [-2.0ex]
   &   $q=4$ &  $q=8$ &  $q=16$&  $q=32$ \\
\hline
\\ [-2.0ex] \hline  \\ [-2.0ex]

%
%
 $n=3$  & 0.2911 (0.2917) &  0.3725 (0.3730) &  0.4299 (0.4302) & 0.4712 (0.4715)\\
 $n=5$  & 0.3977 (0.3977) &  0.4987 (0.4987) &  0.5664 (0.5664) & 0.6131 (0.6131) \\
 $n=10$ & 0.5048 (0.5048) &  0.6207 (0.6207) &  0.6940 (0.6940) & 0.7421 (0.7421) \\
\hline
\end{tabular}
\end{table}




\section{EXIT chart Analysis} \label{sec:exit}

In this section, we analyze the convergence properties of iterative
decoding of WNRMA codes on the QSC using the turbo principle by
means of an EXIT chart analysis \cite{ten01}.

The QSC is characterized by a single parameter $p$, which is the
error probability of the channel. The QSC with error probability $p$
takes a $q$ary symbol at the input and outputs either the unchanged
input symbol, with probability $1-p$, or any of the other $q-1$
symbols, with equal probability, i.e., with probability $p/(q-1)$.
The capacity $C$ (in bits per channel use) of the QSC with error
probability $p$, assuming that $q=2^m$, for some positive integer
$m$, is given by \cite{wei08}
\begin{equation} \label{eq:cap}
C = m - \H_2(p) - p \log_2(q - 1).
\end{equation}

Asymptotically, the normalized
capacity $C/m$ approaches $1-p$ as $m$ tends to infinity, which is
the capacity of a binary erasure channel with erasure probability
$p$.

The EXIT chart for a WNRMA code ensemble can be computed by properly
combining the EXIT functions of the $L$ constituent encoders
$C_0,\ldots,C_{L-1}$ (see Fig.~\ref{fig:encoderWNRMA}) into a single
EXIT function and then plot together in a two-dimensional chart this
EXIT function and the EXIT function of encoder $C_{L}$. Note that
each encoder is followed by a nonbinary RW. However, for notational
simplicity, we assume that the RWs are included in
$C_0,\ldots,C_{L-1}$, i.e., when we speak about the EXIT function of
$C_l$, we are referring to the EXIT function of encoder $C_l$
followed by a RW. Let $I^{C_l}_{e,\u_l}$ and $I^{C_l}_{e,\x_l}$
denote the extrinsic mutual information (MI) generated by decoder
$C^{-1}_l$ on input word $\u_l$ at the input of encoder $C_l$ and on
codeword $\x_l$ at the output of $C_l$, respectively. Likewise, we
define the \emph{a priori} MIs by $I^{C_l}_{a,\u_l}$ and
$I^{C_l}_{a,\x_l}$. Consider the WNRMA encoder with encoders
$C_0,\ldots,C_{L-1}$ as a single encoder and denote it by $C_O$. We
can compute the EXIT functions
\begin{equation}\label{eq:EXITfunctions}
I^{C_O}_{e,\x_{L-1}}=T^{C_O}(I^{C_O}_{a,\x_{L-1}}) \text{ and }
I^{C_L}_{e,\u_L}=T^{C_L}(I^{C_L}_{a,\u_L},p)
\end{equation}
for encoders $C_O$ and $C_L$, respectively, where
$I^{C_O}_{a,\x_{L-1}}=I^{C_L}_{e,\u_L}$ and
$I^{C_L}_{a,\u_L}=I^{C_O}_{e,\x_{L-1}}$. Note  that
$I^{C_O}_{e,\x_{L-1}}$ does not depend on the channel while
$I^{C_L}_{e,\u_L}$ does, since $C_L$ is connected to the channel.

The convergence behavior of WNRMA codes can now be tracked by
displaying in a single plot the two EXIT curves in
(\ref{eq:EXITfunctions}). The iterative decoding of WNRMA codes
processes extrinsic information at the symbol level. Therefore,
nonbinary EXIT chart analysis is required. To compute the EXIT
functions in (\ref{eq:EXITfunctions}) we use the method proposed in
\cite{Sca01} for turbo codes and serially concatenated codes,
generalized to multiple serially concatenated codes.

\begin{table}[!t]
\scriptsize \centering \caption{Convergence thresholds for WNRAA
code ensembles for different values of the repetition factor $n$ and
the field size $q$ on the QSC. The corresponding capacity values
(computed from (\ref{eq:cap}) with $C/ \log_2(q)=1/n$) are given in
the parentheses.} \label{tab:thresholds}
\def\Hline{\noalign{\hrule height 2\arrayrulewidth}}
\vskip -3.0ex 
\begin{tabular}{lcccc}
\Hline \\ [-2.0ex]
   &   $q=4$ &  $q=8$ &  $q=16$&  $q=32$ \\
\hline
\\ [-2.0ex] \hline  \\ [-2.0ex]

 $n=3$  & 0.228 (0.292) &  0.290 (0.373) &  0.335 (0.430) & 0.374 (0.471)\\
 $n=5$  & 0.263 (0.398) &  0.333 (0.499) &  0.382 (0.566) & 0.412 (0.613)\\
 $n=10$ & 0.306 (0.505) &  0.379 (0.621) &  0.424 (0.694) & 0.459 (0.742)\\
\hline
\end{tabular}
\vspace{-2mm}
\end{table}

The convergence thresholds for WNRAA code ensembles predicted by the
EXIT chart analysis (the maximum value of $p$ such that a tunnel
between the two EXIT curves in (\ref{eq:EXITfunctions}) is observed)
are given in Table~\ref{tab:thresholds} for several values of the
repetition factor $n$ and the field size $q$ on a QSC. For
comparison purposes, we also report in Table~\ref{tab:thresholds}
the corresponding capacity values computed from (\ref{eq:cap}). From
Table~\ref{tab:thresholds} it can be observed that for a given $n$
the gap to capacity is similar for different values of $q$. On the
other hand, given $q$, the iterative thresholds of WNRAA code
ensembles are further away from capacity for increasing values of
$n$.

\section{Binary Image of WNRMA Code Ensembles}
\label{sec:BinaryImage}

In this section, we consider the binary image of WNRMA code
ensembles. We derive the average binary WE over the ensemble of
binary images of WNRMA code ensembles, where each nonzero symbol
from GF($q$) is mapped uniformly at random to nonzero binary vectors. We then
compute the $\dmin$ asymptotic growth rates and upper bounds on the  ML thresholds over an
AWGN channel and compare them with
those of binary RMA code ensembles.

Let $C$ be a code of length $N$ over GF($q$), where $q=2^m$, and let
$a^C_h$ be its nonbinary WE. Denote by $\bar{a}^{C,\mathrm{b}}_d$
the average binary WE over the ensemble of binary images of $C$,
where each nonzero symbol from GF($q$) is mapped uniformly at random
to nonzero binary vectors of length $m$, giving the number of
codewords of binary weight $d$. In the following, we will refer to
this WE as the average binary WE of code $C$. Also, denote as before
by $\rho=\frac{h}{N}$ the normalized nonbinary weight. Likewise, we
denote by $\delta=\frac{d}{Nm}$ the normalized binary weight. The
average binary WE of code $C$ can be obtained from the nonbinary WE
of the code as \cite{ElK04}
\begin{equation}\label{eq:bWEfromNbWE}
\begin{split}
&\bar{a}^{C,\mathrm{b}}_{\left \lfloor \delta Nm \right \rfloor} = \\
&\sum_{i= \left \lfloor \delta N \right \rfloor }^{\min(N,\left \lfloor \delta N
m \right \rfloor)} \frac{a^C_i}{\left( 2^m-1 \right)^i} \coef \left( \left(
(1+x)^{m}-1 \right)^i,x^{\left \lfloor \delta N m \right \rfloor } \right)
\end{split}
\end{equation}
where $\coef(p(x),x^i)$ is a shorthand notation for the coefficient of the monomial $x^i$ (the second argument) in the polynomial $p(x)$ (the first argument).

The binary asymptotic spectral shape function of the WNRMA code ensemble
is defined as \cite{Gal63}
\begin{equation} \label{eq:spshapebinary}
        r^{\mathcal{C}_{\mathrm{WNRMA}}}_\mathrm{b}(\delta) = \limsup_{N\longrightarrow\infty}\frac{1}{Nm}\ln\abar^{\mathcal{C}_{\mathrm{WNRMA},{\rm b}}}_{\left \lfloor\delta Nm \right \rfloor}
\end{equation}
where $\abar^{\mathcal{C}_{\mathrm{WNRMA},{\rm b}}}_i$ denotes the average binary WE of the overall ensemble consisting of all possible binary images (obtained by mapping nonzero symbols from GF($q$) to nonzero binary vectors of length $m$) of all codes in the WNRMA code ensemble.

We will make use of the following corollary.
\begin{corollary}[\cite{di06}, Corollary 16 with $d=1$]\label{cor:corollary}
\begin{equation} \notag
\lim_{N\rightarrow \infty}\frac{1}{N} \ln \coef (p(x)^N,x^{N\xi})=\ln
p(\tilde{x})-\xi\ln \tilde{x}
\end{equation}
where $p(x)$ is a polynomial in $x$ and $\tilde{x}$ is the smallest positive solution of
\begin{displaymath}
\left.\frac{\partial\ln p(\mathrm{e}^{s})}{\partial
s}\right|_{s=\ln x}=\xi.
\end{displaymath}
\end{corollary}

Using (\ref{eq:bWEfromNbWE}) and Corollary~\ref{cor:corollary} in
(\ref{eq:spshapebinary}), the binary asymptotic spectral shape
function of the WNRMA code ensemble can be written as
\begin{equation} \label{eq:spshapebinary2}
\begin{split}
&r^{\mathcal{C}_{\mathrm{WNRMA}}}_\mathrm{b}(\delta)\\
&\;\;\;\;= \frac{1}{m}  \sup_{\delta \leq \rho \leq \min(1,m \delta)} \left(  r^{\mathcal{C}_{\mathrm{WNRMA}}}(\rho) -\rho \ln(2^m-1) \right. \\
&\;\;\;\;\;\;\;\;\;\;\;\;\;\;\;\;\;\;\;\;\;\;\;\;\;\;\;\;\;\;\;\;\;\;\;\;+
\rho \ln \left( (1+\tilde{x})^m-1 \right) - \left. m \delta
\ln \tilde{x} \right)
\end{split}
\end{equation}
where $\tilde{x}$ is the smallest positive solution to the
polynomial equation
\begin{equation} \notag
\rho x(1+x)^{m-1} = \delta ((1+x)^m-1)
\end{equation}
which simplifies, using the binomial theorem, to
\begin{equation} \notag
\sum_{j=1}^m \left( 1-\frac{m\delta}{j \rho} \right)
\binom{m-1}{j-1} x^j = 0.
\end{equation}

To analyze the asymptotic behavior of the binary $\dmin$  of WNRMA
code ensembles, we must solve the optimization problem in
(\ref{eq:spshapebinary2}), similarly to the nonbinary case. Notice
that since the nonbinary $\dmin$ of WNRMA code ensembles grows
linearly with the block length, see Theorem~\ref{th:rma_growth2}, it follows that the $\dmin$ of its
binary image also grows linearly with the block length. In
Table~\ref{tab:crossingsL2_binary}, we give the binary $\dmin$ growth rate
coefficient $\delta_0$ of WNRAA code ensembles for several values of
$n$ and $q$. As a comparison, we also report the values for the
binary RAA code ensemble ($q=2$). It is observed that WNRAA code
ensembles achieve higher growth rates than RAA code ensembles. For $q=32$
the growth rates are very close to the GVB. The asymptotic binary $\dmin$ growth rates for
WNRAAA and RAAA code ensembles are reported in
Table~\ref{tab:crossingsL3_binary}.

\begin{table}[!t]
\setlength{\tabcolsep}{4.0pt}
\scriptsize \centering \caption{Binary image growth rate coefficient
$\delta_0$ of WNRAA codes for different values of the repetition
factor $n$ and the field size $q$. The corresponding growth rates
from the asymptotic binary GVB are given in the last column.}
\label{tab:crossingsL2_binary}
\def\Hline{\noalign{\hrule height 2\arrayrulewidth}}
\vskip -3.0ex 
\begin{tabular}{llccccc}
\Hline \\ [-2.0ex]
   &   $q=2$ & $q=4$ &  $q=8$ &  $q=16$&  $q=32$ & GVB \\
\hline
\\ [-2.0ex] \hline  \\ [-2.0ex]

%
%
%
 $n=3$  & 0.1323 \cite{fan09,all07} & 0.1496 & 0.1608 & 0.1675 & 0.1712 & 0.1740\\
 $n=5$  & 0.2286 \cite{fan09,all07} & 0.2380 & 0.2416 & 0.2427 & 0.2429 & 0.2430 \\
 $n=10$ & 0.3133 \cite{fan09} & 0.3155 & 0.3159 & 0.3160 & 0.3160 & 0.3160 \\
\hline
\end{tabular}
\end{table}
\begin{table}[!t]
\setlength{\tabcolsep}{4.0pt}
\scriptsize \centering \caption{Binary image growth rate coefficient
$\delta_0$ of WNRAAA codes for different values of the repetition
factor $n$ and the field size $q$. The corresponding growth rates
from the asymptotic binary GVB are given in the last column.}
\label{tab:crossingsL3_binary}
\def\Hline{\noalign{\hrule height 2\arrayrulewidth}}
\vskip -3.0ex 
\begin{tabular}{llccccc}
\Hline \\ [-2.0ex]
   &   $q=2$ & $q=4$ &  $q=8$ &  $q=16$&  $q=32$ & GVB\\
\hline
\\ [-2.0ex] \hline  \\ [-2.0ex]

%
%
%
 $n=3$  & 0.1731 \cite{fan09,all07} & 0.1738 & 0.1739 & 0.1739 & 0.1740  & 0.1740 \\
 $n=5$  & 0.2430 \cite{fan09,all07} & 0.2430 & 0.2430 & 0.2430 & 0.2430  & 0.2430 \\
 $n=10$ & 0.3160 \cite{fan09} & 0.3160 & 0.3160 & 0.3160 & 0.3160  & 0.3160 \\
\hline
\end{tabular}
\end{table}

\subsection{Threshold Under ML Decoding}

The asymptotic spectral shape function of a code ensemble can also
be used to derive a threshold under ML decoding.
An upper bound on the ML decoding threshold of a code ensemble on the AWGN channel, due
to Divsalar \cite{div99}, is given by
\begin{equation}\label{eq:MLbound}
\left( \frac{E_b}{N_0} \right)_{\rm ML, threshold} \leq \frac{1}{R}
\cdot \max_{0 \leq \rho \leq 1} \left[ \frac{(1-{\rm e}^{-2
r(\rho)})(1-\rho)}{2 \rho} \right]
\end{equation}
where $R$ is the code rate, $r(\rho)$ is the asymptotic spectral
shape function, $E_b/N_0$ denotes the signal-to-noise ratio (SNR),
and $(E_b/N_0)_{\rm ML, threshold}$ is the ML decoding threshold. We
computed the upper bound on the ML decoding threshold in
(\ref{eq:MLbound}) numerically for the binary image of both WNRAA and WNRAAA code
ensembles for several values of $n$ and $q$. The results are given
in Tables~\ref{table:ThresholdsMLRAA} and
\ref{table:ThresholdsMLRAAA}, respectively. For comparison purposes,
we also report in the table the binary-input AWGN Shannon limit. All
codes perform within $0.05$ dB from capacity.


\begin{table*}[!t]
\scriptsize \centering \caption{Upper bounds on the ML decoding threshold of the binary image of
WNRAA codes based on Divsalar's bound in \cite{div99}.}\label{table:ThresholdsMLRAA}
\def\Hline{\noalign{\hrule height 2\arrayrulewidth}}
\vskip -3.0ex 
\begin{tabular}{lcccccc}
\Hline \\ [-2.0ex]
   &   $q=2$ &  $q=4$ &  $q=8$ &  $q=16$&  $q=32$ & Capacity\\
\hline
\\ [-2.0ex] \hline  \\ [-2.0ex]

 $n=3$  & -0.437 & -0.449 dB &  -0.453 dB & -0.453 dB & -0.453 dB & -0.495 dB \\
 $n=5$  & -0.952 & -0.953 dB &  -0953  dB & -0.953 dB & -0.953 dB & -0.964 dB\\
 $n=10$ & -1.284 & -1.284 dB &  -1.284 dB & -1.284 dB & -1.284 dB & -1.286 dB\\
\hline
\end{tabular}
\end{table*}

\begin{table*}[!t]
\scriptsize \centering \caption{Upper bounds on the ML decoding threshold of the binary image of
WNRAAA codes based on Divsalar's bound in \cite{div99}.}\label{table:ThresholdsMLRAAA}
\def\Hline{\noalign{\hrule height 2\arrayrulewidth}}
\vskip -3.0ex 
\begin{tabular}{lcccccc}
\Hline \\ [-2.0ex]
   &   $q=2$ &  $q=4$ &  $q=8$ &  $q=16$&  $q=32$ & Capacity\\
\hline
\\ [-2.0ex] \hline  \\ [-2.0ex]

 $n=3$  & -0.453 & -0.453 dB &  -0.453 dB & -0.453 dB & -0.453 dB & -0.495 dB \\
 $n=5$  & -0.953 & -0.953 dB &  -0953  dB & -0.953 dB & -0.953 dB & -0.964 dB\\
 $n=10$ & -1.284 & -1.284 dB &  -1.284 dB & -1.284 dB & -1.284 dB & -1.286 dB\\
\hline
\end{tabular}
\vspace{-2mm}
\end{table*}

\subsection{Convergence Thresholds Under Iterative Decoding}

In Table~\ref{tab:thresholdsBinary}, we report the iterative
convergence thresholds for the binary image of WNRAA code ensembles
on the AWGN channel for repeat factor $n=3$ and $n=5$.
Unfortunately, while the ML thresholds improve with $q$ and get
closer to the capacity, the iterative convergence thresholds get
worse with increasing values of $q$. We remark that if we remove the
nonbinary weighters, the iterative decoding thresholds will improve,
and they will be slightly better than those of binary RMA codes with
the same repetition factor $n$.

However, for the QSC, the iterative decoding thresholds will be the
same with and without the nonbinary weighter (random or fixed). This
can be explained by the following argument. Let the symbol-wise
log-likelihood ratio (LLR) for the $i$th received symbol $r_i$,
$i=1,\dots,N$,  be defined as the length-$(q-1)$ vector
\begin{displaymath}
\left( \ln \left( \frac{P(r_i|1)}{P(r_i|0)} \right),\dots,\ln \left(
\frac{P(r_i|q-1)}{P(r_i|0)} \right) \right)
\end{displaymath}
where $P(r_i | x_i)$ is the probability of receiving $r_i$ when
$x_i$ is transmitted over the QSC. Under the assumption that we
transmit  the all-zero codeword, the symbol-wise LLR vectors will be
of the form
\begin{displaymath}
\left( \ln \left( \frac{p}{(1-p)(q-1)} \right),\dots,  \ln \left(
\frac{p}{(1-p)(q-1)} \right) \right)
\end{displaymath}
when we receive a zero (with probability $1-p$), or
\begin{equation} \label{eq:symbolLLR}
\left( \overbrace{0,\dots,0}^{j-1},\ln \left( \frac{(1-p)(q-1)}{p}
\right),\overbrace{0,\dots,0}^{q-1-j} \right)
\end{equation}
when $r_i=j$, $j > 0$. The probability of having such an LLR vector
is $p/(q-1)$, independent of $j$ (and of $i$). Since the nonbinary
weighter will map a zero to a zero, the probability distribution of
the LLR vector is \emph{preserved} by the nonbinary weighter.
Furthermore, due to properties of the rate-$1$ nonbinary accumulator
(with no trellis termination), all symbol-wise LLR vectors of the
form in (\ref{eq:symbolLLR}) have the same probability, independent
of $j$, even at the input of the $L$th nonbinary accumulator (after
decoding on the $L$th nonbinary accumulator trellis). Due to this
property, the nonbinary weighter at stage $L-1$ (after $C_{L-1}$)
will also \emph{preserve} the probability distribution of the
symbol-wise LLR vector. The result follows by induction on the
number of accumulators in the WNRMA code ensemble.

Finally, we remark that the above argument does not apply to the
binary image of WNRMA code ensembles transmitted over the
binary-input AWGN channel, since the \emph{symmetry} of the QSC is
lost.

\begin{table}[!t]
\scriptsize \centering \caption{Convergence thresholds for the
binary image of WNRAA code ensembles for different values of the
repetition factor $n$ and the field size $q$ on the AWGN channel.}
\label{tab:thresholdsBinary}
\def\Hline{\noalign{\hrule height 2\arrayrulewidth}}
\vskip -3.0ex 
\begin{tabular}{lccccc}
\Hline \\ [-2.0ex]
   &   $q=2$ &  $q=4$ &  $q=8$ &  $q=16$&  $q=32$ \\
\hline
\\ [-2.0ex] \hline  \\ [-2.0ex]

 $n=3$  & 1.68 dB &  1.94 dB &  2.23 dB & 2.45 dB & 2.66 dB \\
 $n=5$  & 2.77 dB &  3.10 dB &  3.46 dB & 3.76 dB & 4.07 dB \\
\hline
\end{tabular}
\vspace{-2mm}
\end{table}

\section{Conclusion}
\label{sec:conclusion}

In this paper, we analyzed the symbol-wise minimum distance
properties of WNRMA code ensembles, where each encoder is followed
by a nonbinary random weighter. We derived an exact closed-form
expression for the IOWE of nonbinary accumulators. Based on that, we
derived the ensemble-average WE of WNRMA code ensembles and analyzed
its asymptotic behavior. Furthermore, we formally proved that the
symbol-wise minimum distance of WNRMA code ensembles asymptotically
grows linearly with the block length when $L \geq 3$ and $n \geq 2$,
and $L=2$ and $n \geq 3$, for all powers of primes $q \geq 3$
considered. The asymptotic growth rate coefficient of the minimum distance of
WNRAA and WNRAAA code ensembles for different values of the
repetition factor $n$ and the field size $q$ were also computed. The
asymptotic growth rates are very close to the GVB when $q$ is large,
but not too large. We also considered EXIT charts and analyzed the
iterative convergence behavior of WNRMA code ensembles on the QSC.
Finally, we considered the binary image of WNRMA code ensembles. We
computed the asymptotic growth rates of their minimum distance and
compared them with those of binary RMA code ensembles. It is shown
that WNRMA code ensembles achieve higher $\dmin$ growth rates than
their binary counterparts, and they get close to the GVB when $q$
and $n$ grow. Upper bounds on the ML decoding thresholds of the
binary image of WNRMA code ensembles on the AWGN channel were also
computed, and it was shown that WNRMA code ensembles perform very
close to capacity under ML decoding. Unfortunately, iterative
decoding is not able to fully exploit the performance of WNRMA code
ensembles on the binary-input AWGN channel. In other words, WNRMA
codes are excellent codes, but there is no efficient decoding
algorithm to decode them close-to-ML. However, on the QSC, the
iterative decoding performance is closer to capacity.

\section*{Appendix~A\\Numerical Evaluation of the Asymptotic Spectral Shape Function in (\ref{eq:spshape2})}

The optimal value for the objective function in (\ref{eq:objWNRMA}) can be either at a stationary point or at the boundary. To find the stationary points, we compute the partial derivatives with respect to $\beta_0,\beta_1,\dots,\beta_{L-1}$ and $\gamma_1,\dots,\gamma_L$.  Setting the partial derivative with respect to $\beta_0$ equal to zero gives the equation 
\begin{equation} \label{eq:beta0}
\left( \frac{(q-1) (1-\beta_0)}{\beta_0} \right)^{1/n-1} \cdot
\left( \frac{\beta_1-\beta_0+\gamma_1}{\beta_0-2\gamma_1} \right)  =
\frac{1}{q-2}.
\end{equation}
Setting the partial derivative with respect to $\gamma_l$, $l=1,\dots,L$, equal to zero results in the equation
\begin{equation} \label{eq:gamma}
\begin{split}
&(q-2)^2 \left( \beta_l-\beta_{l-1}+\gamma_l \right) \gamma_l^2\\
&~~~~= (q-1) \left( \beta_{l-1}-2\gamma_l \right)^2 \left( 1-\beta_l-\gamma_l \right).
\end{split}
\end{equation}
Finally, setting the partial derivative with respect to $\beta_l$, $l=1,\dots,L-1$, equal to zero results in the equation
\begin{equation} \label{eq:beta}
\frac{\beta_l^2 (1-\beta_l-\gamma_l) (\beta_{l+1}-\beta_l+\gamma_{l+1}) }{(1-\beta_l)^2(\beta_l-\beta_{l-1}+\gamma_l) (\beta_l-2\gamma_{l+1})} = \frac{q-1}{q-2}.
\end{equation}

To determine a solution to the above set of equations, we choose the
following strategy. First treat $\beta_0$ as a free parameter. Then, from (\ref{eq:beta0}), solve for $\beta_1$ as a (linear) function of $\gamma_1$ and insert the resulting expression into (\ref{eq:gamma}) for $l=1$. The resulting third order equation can now be solved for $\gamma_1$. Using (\ref{eq:beta}) (with $l=1$), we can find $\beta_2$ as a (linear) function of $\gamma_2$ which we again can insert into (\ref{eq:gamma}) for $l=2$. The resulting third order equation can then be solved for $\gamma_2$. Continuing like this, we can find the remaining values for $\beta_l$ and $\gamma_l$.

In general, we must also consider all combinations of boundary conditions in a systematic way, since the optimum value may be at the boundary. Details are omitted for brevity. 

\section*{Appendix~B \\ Proof of Theorem~\ref{th:rma_growth}}
\label{Ap:AppendixC}

The proof of Theorem~\ref{th:rma_growth} follows closely the proof
of Theorem 6 in \cite{GraRos09}, which is inspired by the proof of
Theorem 6 (or Theorem 9)\footnote{In \cite{fan09}, Theorems 6 and 9 are the same result.} in \cite{fan09} and the asymptotic techniques devised in
\cite{jin02}.  We start by proving Lemma~\ref{lem:lemma2} stated
below. The lemma is proved by induction on $L$.

\begin{lemma}\label{lem:lemma2}
Let $\{h_N\}_{N \in \mathbb{N}}$ be a sequence of integers such that
for any arbitrary $\eta > 0$
\begin{displaymath}
\lim_{N \longrightarrow \infty} \frac{h_N}{N^{\eta}}=0 \text{ and }
\lim_{N \longrightarrow \infty} \frac{\ln h_N}{h_N}=0.
\end{displaymath}
Then,
\begin{displaymath}
\sum_{h=1}^{h_N} \bar{a}_h^{\mathcal{C}_{\rm WNRMA}} = O \left(
N^{1-\sum_{i=1}^L \left\lceil \frac{n}{2^i} \right\rceil +\eta}
\right)
\end{displaymath}
where $L$ is the number of accumulators.
\end{lemma}

\begin{IEEEproof}
We prove the lemma by induction on the number of accumulators $L$.
Consider first the case of $L=1$. We have
\begin{displaymath}
\begin{split}
\sum_{h=1}^{h_N} \bar{a}^{\mathcal{C}_{\rm WNRA}}_h &= \sum_{w=1}^{2
h_N/n} \binom{N/n}{w } \left(q-1\right)^w \\
&\times \sum_{h=1}^{h_N} \frac{
                      \sum_{k=1}^{\left\lfloor\frac{nw}{2}\right\rfloor}{N-h \choose k}{h-1 \choose k-1}{h-k \choose nw-2k} \left( q-1 \right)^{k-nw}
}{{N \choose nw} \left( q-2 \right)^{2k-nw} }    \\
&\leq \sum_{w=1}^{2 h_N/n} N^{w -\left\lceil \frac{nw}{2} \right\rceil} g(w,N)
\sum_{h=1}^{h_N} h^{nw+\left\lfloor \frac{nw}{2} \right\rfloor-3}
\end{split}
\end{displaymath}
where 
\begin{displaymath}
\begin{split}
g(w,N) &=\frac{(nw)!\mathrm{e}^{nw+w-1}\varphi_N(nw-1)}{n^ww^w} \\
&~~\;\; \times \sum_{k=1}^{\left\lfloor\frac{nw}{2}\right\rfloor}
\frac{(nw-2k)^{2k-nw}(q-1)^{k-nw+w}}{k^k(k-1)^{k-1} (q-2)^{2k-nw}}
\end{split}
\end{displaymath}
and we have used the Stirling's approximation ${n \choose k }\leq \left( \frac{n \mathrm{e}}{k} \right)^k$ and the fact that $\prod_{i=0}^l \left(N-i \right) \geq \frac{N^{l+1}}{\phi_N(l)}$, with $\phi_{\lambda}(l) = \exp \left( \frac{l(l+1)}{2 \lambda} \right)$.
Also, note that the upper bound of $2h_N/n$ in the summation over $w$ is
due to the binomial $\binom{h-k}{nw-2k}$. In more detail, $h-k \geq
nw-2k$, from which it follows that $nw \leq h+k \leq h+\left\lfloor nw/2
\right\rfloor$, which implies that $w \leq 2h/n$. Now, it follows that
\begin{displaymath}
\begin{split}
\sum_{h=1}^{h_N} \bar{a}^{\mathcal{C}_{\rm RA}}_h &\leq
\sum_{w=1}^{2 h_N/n} N^{w -\left\lceil \frac{nw}{2} \right\rceil} g(w,N) h_N^{nw+\left\lfloor \frac{nw}{2} \right\rfloor-2}  \\
& \leq \frac{2h_N}{n} \max_{1 \leq w \leq 2h_N/n} N^{w -\left\lceil \frac{nw}{2} \right\rceil} g(w,N) h_N^{nw+\left\lfloor \frac{nw}{2} \right\rfloor-2} \\
&\leq \frac{2}{n} N^{1 -\left\lceil \frac{n}{2} \right\rceil+\eta} g(1,N)
h_N^{n+\left\lfloor \frac{n}{2} \right\rfloor-1} \\
&= O \left( N^{1-\left\lceil
n/2 \right\rceil+\eta} \right)
\end{split}
\end{displaymath}
for large enough $N$ and for all $\eta > 0$. Note that for large
enough $N$, $N^{w -\left\lceil \frac{nw}{2} \right\rceil}$ dominates $g(w,N)
h_N^{nw+\left\lfloor \frac{nw}{2} \right\rfloor-2}$, due to the conditions on
$h_N$ stated in the lemma.
Now, assume that the statement of the lemma is true for the case of
$L-1$. We get
\begin{displaymath}
\begin{split}
&\sum_{h=1}^{h_N} \bar{a}^{\mathcal{C}_{{\rm WNRMA}(L)}}_h \\
&~~~=\sum_{w=\left \lceil \frac{n}{2^{L-1}} \right \rceil}^{2h_N} \bar{a}^{\mathcal{C}_{{\rm WNRMA}(L-1)}}_w  \\
&~~~~~\;\; \times \sum_{h=1}^{h_N}\sum_{k=1}^{\left\lfloor\frac{w}{2}\right\rfloor} \frac{{N-h \choose k}{h-1 \choose k-1}{h-k \choose w-2k} \left(q-1\right)^{k-w}}{{N \choose w} \left(q-2 \right)^{2k-w}} \\
&~~~\leq \sum_{w=\left \lceil \frac{n}{2^{L-1}} \right \rceil}^{2h_N}
\bar{a}^{\mathcal{C}_{{\rm WNRMA}(L-1)}}_w  N^{-\left\lceil \frac{w}{2} \right\rceil} g'(w,N) \sum_{h=1}^{h_N}
 h^{w+\left\lfloor \frac{w}{2} \right\rfloor-3}
\end{split}
\end{displaymath}
where
\begin{displaymath}
\begin{split}
g'(w,N) &=
(w)!\mathrm{e}^{w-1}\varphi_N(w-1) \\
&~~\;\; \times \sum_{k=1}^{\left\lfloor\frac{w}{2}\right\rfloor}\frac{(w-2k)^{2k-w}(q-1)^{k-w} }{k^k(k-1)^{k-1} (q-2)^{2k-w}}
\end{split}
\end{displaymath}
and $\mathcal{C}_{{\rm WNRMA}(l)}$  denotes the WNRMA code ensemble
with $l$ accumulators.
Note that the lower bound of $\left \lceil n / 2^{L-1} \right
\rceil$ in the summation over $w$ is due to the fact that the output
size $h$ of an accumulator is at least $\left\lceil w/2
\right\rceil$, where $w$ is the input size. This is due to the
binomial $\binom{h-k}{w-2k}$ and the upper bound of $\left\lfloor
w/2 \right\rfloor$ in the summation over $k$.
It follows that
\begin{displaymath}
\begin{split}
&\sum_{h=1}^{h_N} \bar{a}^{\mathcal{C}_{{\rm WNRMA}(L)}}_h \\
&~~~~\leq \sum_{w= \left \lceil \frac{n}{2^{L-1}} \right \rceil }^{2h_N} \bar{a}^{\mathcal{C}_{{\rm WNRMA}(L-1)}}_w  N^{-\left\lceil \frac{w}{2} \right\rceil}  g'(w,N) h_N^{w+\left\lfloor \frac{w}{2} \right\rfloor-2}  \\
&~~~~\leq O \left( N^{1-\sum_{i=1}^{L-1} \left\lceil \frac{n}{2^i} \right\rceil +\eta} \right) \\
&~~~~\;\;\;\;\times \max_{ \left \lceil \frac{n}{2^{L-1}} \right \rceil \leq w \leq 2h_N} N^{-\left\lceil \frac{w}{2} \right\rceil}  g'(w,N) h_N^{w+\left\lfloor \frac{w}{2} \right\rfloor-2}  \\
&~~~~= O \left( N^{1-\sum_{i=1}^{L} \left\lceil \frac{n}{2^i} \right\rceil
+\eta} \right)
\end{split}
\end{displaymath}
for large enough $N$ and for all $\eta > 0$. Above, we used the induction hypothesis in the second inequality. Also, note that for large enough $N$, $N^{-\left\lceil \frac{w}{2} \right\rceil}$
dominates $g'(w,N) h_N^{w+\left\lfloor \frac{w}{2} \right\rfloor-2}$, due to
the conditions on $h_N$ stated in the lemma.
\end{IEEEproof}

\begin{lemma}\label{lem:lemma3}
Let $r^{\mathcal{C}_{\rm WNRMA}}(\rho;N)$ denote the $N$th  spectral shape function of the WNRMA code ensemble, defined
as $r^{\mathcal{C}_{\rm WNRMA}}(\rho;N) = \frac{1}{N} \ln
\bar{a}^{\mathcal{C}_{\rm WNRMA}}_{\left\lfloor \rho N \right\rfloor}$. Then,
\begin{displaymath}
r^{\mathcal{C}_{\rm WNRMA}}(\rho;N) \leq \frac{2L \ln(N+1)}{N} + r^{\mathcal{C}_{\rm WNRMA}}(\rho).
\end{displaymath}
\end{lemma}
\begin{IEEEproof}
The proof of the lemma relies on the function $\psi(u,\rho)$,
defined in (\ref{eq:f}). In particular, the proof of the lemma is by induction on $L$,
following the same arguments as in the proof of Lemma 5 in \cite{fan09}, 
and is therefore omitted for brevity.
\end{IEEEproof}

The final part of the proof of \cite[Theorem 9]{fan09} is
also very general, and it can easily be extended to the case of WNRMA codes. In fact, the rest of the proof only relies on the following
properties of $\psi(u,\rho)$.
\begin{enumerate}
\item $\psi(u,\rho)$ is continuous;
\item $\psi(u,\rho)$, for fixed $u$, is strictly increasing in $\rho < 1/2$;
\item $\frac{\psi(u,\rho)}{u}$, for fixed $\rho$, is decreasing in $u$; and
\item $\lim_{u \longrightarrow 0} \frac{\psi(u,\rho)}{u} < 0~\forall \rho<(q-1)/q$.
\end{enumerate}
Finally, by using Lemmas~\ref{lem:lemma2} and~\ref{lem:lemma3} and the properties
above, Theorem~\ref{th:rma_growth} is proved following the same
arguments as in the proof of \cite[Theorem 9]{fan09}. Below, we will prove Properties 2 to 4. Property 1 follows from the definition of $\psi(u,\rho)$ given in (\ref{eq:f}).

\subsection{Proof of Property 2} \label{sec:property2}

To prove Property 2, we compute the partial derivative of $\psi(u,\rho)$ with respect to $\rho$. Let the optimum value of $\gamma$ (as a function of $\rho$) when solving the optimization problem in (\ref{eq:f}) be denoted by $\hat{\gamma}_{\rho} = \hat{\gamma}(\rho)$ and its derivative with respect to $\rho$ as $\hat{\gamma}_{\rho}' = \hat{\gamma}'(\rho)$. Now,
\begin{equation} \label{eq:derivativerho}
\begin{split}
\frac{\partial \psi(u,\rho)}{\partial \rho} &= \H \left( \frac{\hat{\gamma}_{\rho}}{\rho} \right) + \frac{\rho \hat{\gamma}_{\rho}' -\hat{\gamma}_{\rho}}{\rho} \ln \left( \frac{\rho-\hat{\gamma}_{\rho}}{\hat{\gamma}_{\rho}} \right) \\
&\;\;\;\; -\H \left( \frac{\hat{\gamma}_{\rho}}{1-\rho} \right)+\frac{ \hat{\gamma}_{\rho}'(1-\rho) + \hat{\gamma}_{\rho}}{1-\rho} \ln \left( \frac{1-\rho-\hat{\gamma}_{\rho}}{\hat{\gamma}_{\rho}} \right) \\
&\;\;\;\; +(1-\hat{\gamma}_{\rho}') \H \left( \frac{u-2 \hat{\gamma}_{\rho}}{\rho - \hat{\gamma}_{\rho}} \right) \\
&\;\;\;\; + \frac{\hat{\gamma}_{\rho}'(u-2\rho)-u+2\hat{\gamma}_{\rho}}{\rho-\hat{\gamma}_{\rho}} \ln \left( \frac{\rho+\hat{\gamma}_{\rho}-u}{u-2\hat{\gamma}_{\rho}} \right) \\
&\;\;\;\; + \hat{\gamma}_{\rho}' \ln \left(q-1 \right)-2\hat{\gamma}_{\rho}' \ln \left(q-2 \right)\\
&= \H \left( \frac{\hat{\gamma}_{\rho}}{\rho} \right) - \frac{\hat{\gamma}_{\rho}}{\rho} \ln \left( \frac{\rho-\hat{\gamma}_{\rho}}{\hat{\gamma}_{\rho}} \right)
-\H \left( \frac{\hat{\gamma}_{\rho}}{1-\rho} \right)\\
&\;\;\;\; +\frac{ \hat{\gamma}_{\rho}}{1-\rho} \ln \left( \frac{1-\rho-\hat{\gamma}_{\rho}}{\hat{\gamma}_{\rho}} \right) + \H \left( \frac{u-2 \hat{\gamma}_{\rho}}{\rho - \hat{\gamma}_{\rho}} \right) \\
&\;\;\;\; - \frac{u-2\hat{\gamma}_{\rho}}{\rho-\hat{\gamma}_{\rho}} \ln \left( \frac{\rho+\hat{\gamma}_{\rho}-u}{u-2\hat{\gamma}_{\rho}} \right) \\
%
%
&\;\;\;\; +\hat{\gamma}_{\rho}' \left[ \ln \left( \frac{\rho-\hat{\gamma}_{\rho}}{\hat{\gamma}_{\rho}} \right) + \ln \left( \frac{1-\rho-\hat{\gamma}_{\rho}}{\hat{\gamma}_{\rho}} \right) \right. \\
&\;\;\;\;\;\;\;\;\left. - \H \left( \frac{u-2\hat{\gamma}_{\rho}}{\rho-\hat{\gamma}_{\rho}} \right) + \frac{u-2\rho}{\rho-\hat{\gamma}_{\rho}} \ln \left( \frac{\rho+\hat{\gamma}_{\rho}-u}{u-2\hat{\gamma}_{\rho}} \right) \right. \\
&\;\;\;\;\;\;\;\;\left. + \ln (q-1)-2 \ln (q-2) \right].
%
%
%
%
%
%
%
%
%
\end{split}
\end{equation}
Since $\tilde{\gamma}_{\rho}$ is a solution to the optimization problem in (\ref{eq:f}), it follows that $\tilde{\gamma}_{\rho}$ is a solution to the equation
\begin{equation} \label{eq:derivativegamma}
\begin{split}
&\ln \left(\frac{\rho-\gamma}{\gamma} \right) + \ln \left(\frac{1-\rho-\gamma}{{\gamma}} \right) - \H \left( \frac{u-2{\gamma}}{\rho-{\gamma}} \right) \\
&+ \frac{u-2\rho}{\rho-{\gamma}} \ln \left( \frac{\rho+{\gamma}-u}{u-2{\gamma}} \right)
+ \ln(q-1) -2\ln(q-2) = 0
\end{split}
\end{equation}
which is obtained by taking the partial derivative with respect to
$\gamma$ of the objective function in (\ref{eq:f}) and setting it
equal to zero. Substituting (\ref{eq:derivativegamma}) (with
$\gamma=\hat{\gamma}_{\rho}$) into (\ref{eq:derivativerho}), we get
\begin{equation} \notag 
\begin{split}
\frac{\partial \psi(u,\rho)}{\partial \rho} &= \H \left( \frac{\hat{\gamma}_{\rho}}{\rho} \right) - \frac{\hat{\gamma}_{\rho}}{\rho} \ln \left( \frac{\rho-\hat{\gamma}_{\rho}}{\hat{\gamma}_{\rho}} \right)
-\H \left( \frac{\hat{\gamma}_{\rho}}{1-\rho} \right)\\
&\;\;\;\; +\frac{ \hat{\gamma}_{\rho}}{1-\rho} \ln \left( \frac{1-\rho-\hat{\gamma}_{\rho}}{\hat{\gamma}_{\rho}} \right) + \H \left( \frac{u-2 \hat{\gamma}_{\rho}}{\rho - \hat{\gamma}_{\rho}} \right) \\
&\;\;\;\; - \frac{u-2\hat{\gamma}_{\rho}}{\rho-\hat{\gamma}_{\rho}} \ln \left( \frac{\rho+\hat{\gamma}_{\rho}-u}{u-2\hat{\gamma}_{\rho}} \right) \\
&= \ln \left( \frac{\rho(1-\rho-\hat{\gamma}_{\rho})}{(1-\rho) (\rho+\hat{\gamma}_{\rho}-u)} \right)
\end{split}
\end{equation}
where the last equality follows from straightforward algebraic manipulations. Now, setting
\begin{equation} \label{eq:10}
\frac{\rho(1-\rho-\hat{\gamma}_{\rho})}{(1-\rho) (\rho+\hat{\gamma}_{\rho}-u)} \leq 1
\end{equation}
gives $\rho \geq 1-\hat{\gamma}_{\rho}/u \geq 1/2$, since $\hat{\gamma}_{\rho} \leq u/2$ (see (\ref{eq:f})) and the denominator in (\ref{eq:10}) is nonnegative (since $\hat{\gamma}_{\rho} > u-\rho$, see (\ref{eq:f})), from which it follows that, for fixed $u$, $\psi(u,\rho)$ is strictly increasing in $\rho$ for $\rho < 1/2$.


\subsection{Proof of Property 3} \label{sec:property3}

To prove Property 3, we compute the partial derivative of $\frac{\psi(u,\rho)}{u}$ with respect to $u$. We get
\begin{displaymath}
\frac{\partial}{\partial u} \left( \frac{\psi(u,\rho)}{u} \right) = \frac{1}{u^2} \left(u \frac{\partial \psi(u,\rho)}{\partial u}-\psi(u,\rho) \right).
\end{displaymath}
Let the optimum value of $\gamma$ (as a function of $u$) when solving the optimization problem in (\ref{eq:f}) be denoted by $\tilde{\gamma}_u = \tilde{\gamma}(u)$ and its derivative with respect to $u$ as $\tilde{\gamma}_u' = \tilde{\gamma}'(u)$. Now,
\begin{equation} \label{eq:derivativepsi}
\begin{split}
\frac{\partial \psi(u,\rho)}{\partial u} &= -\ln \left( \frac{1-u}{u} \right)+\tilde{\gamma}_u' \ln \left(\frac{\rho-\tilde{\gamma}_u}{\tilde{\gamma}_u} \right) \\
&\;\;\;\;+ \tilde{\gamma}_u' \ln \left(\frac{1-\rho-\tilde{\gamma}_u}{\tilde{\gamma}_u} \right) - \tilde{\gamma}_u' \H \left( \frac{u-2\tilde{\gamma}_u}{\rho-\tilde{\gamma}_u} \right) \\
&\;\;\;\;+ \frac{\rho-\tilde{\gamma}_u-2\rho \tilde{\gamma}_u' + u \tilde{\gamma}_u'}{\rho-\tilde{\gamma}_u} \ln \left( \frac{\rho+\tilde{\gamma}_u-u}{u-2\tilde{\gamma}_u} \right) \\
&\;\;\;\;+ (\tilde{\gamma}_u'-1) \ln(q-1) +(1-2\tilde{\gamma}_u') \ln(q-2) \\
&= -\ln \left( \frac{1-u}{u} \right)
+  \ln \left( \frac{\rho+\tilde{\gamma}_u-u}{u-2\tilde{\gamma}_u} \right) \\
&\;\;\;\;- \ln(q-1) +\ln(q-2) \\
&\;\;\;\;+\tilde{\gamma}_u' \left[ \ln \left(\frac{\rho-\tilde{\gamma}_u}{\tilde{\gamma}_u} \right)
+ \ln \left(\frac{1-\rho-\tilde{\gamma}_u}{\tilde{\gamma}_u} \right) \right. \\
&\;\;\;\;\;\;\;\;- \left.  \H \left( \frac{u-2\tilde{\gamma}_u}{\rho-\tilde{\gamma}_u} \right) +\frac{u -2\rho}{\rho-\tilde{\gamma}_u} \ln \left( \frac{\rho+\tilde{\gamma}_u-u}{u-2\tilde{\gamma}_u} \right) \right. \\
&\;\;\;\;\;\;\;\;+ \left. \ln(q-1) -2 \ln(q-2) \right].
\end{split}
\end{equation}
Since $\tilde{\gamma}_u$ is a solution to the optimization problem in (\ref{eq:f}), it follows that $\tilde{\gamma}_u$ is a solution to (\ref{eq:derivativegamma}).
Substituting (\ref{eq:derivativegamma}) (with $\gamma=\tilde{\gamma}_u$) into (\ref{eq:derivativepsi}), we get
\begin{equation} \label{eq:12}
\begin{split}
\frac{\partial \psi(u,\rho)}{\partial u} &= -\ln \left( \frac{1-u}{u} \right)
+  \ln \left( \frac{\rho+\tilde{\gamma}_u-u}{u-2\tilde{\gamma}_u} \right) \\
&\;\;\;\;- \ln(q-1) +\ln(q-2) \\
\end{split}
\end{equation}
from which it follows that
\begin{equation} \label{eq:derivativepsi1}
\begin{split}
\frac{\partial}{\partial u} \left( \frac{\psi(u,\rho)}{u} \right) &= \frac{1}{u^2} \left( -\ln(1-u) + 2\tilde{\gamma}_u \ln(\tilde{\gamma}_u)-\rho \ln(\rho) \right. \\
&\;\;\;\;- \left. (1-\rho) \ln(1-\rho) \right. \\
&\;\;\;\;+ \left. (1-\rho-\tilde{\gamma}_u) \ln (1-\rho-\tilde{\gamma}_u ) \right. \\
&\;\;\;\;- \left. 2\tilde{\gamma}_u \ln(u-2\tilde{\gamma}_u) \right. \\
&\;\;\;\;+ \left. (\rho+\tilde{\gamma}_u) \ln(\rho+\tilde{\gamma}_u-u) \right.\\
&\;\;\;\;- \left. \tilde{\gamma}_u (\ln(q-1)-2 \ln(q-2) \right).
\end{split}
\end{equation}
Note that the equation in (\ref{eq:derivativegamma}) can be simplified to
\begin{equation} \label{eq:derivativegamma1}
\ln \left( \frac{\gamma^2}{(u-2\gamma)^2} \right) = \ln \left( \frac{(q-1)(1-\rho-\gamma)}{(q-2)^2(\rho+\gamma-u)} \right).
\end{equation}
Substituting (\ref{eq:derivativegamma1}) into (\ref{eq:derivativepsi1}), we get
\begin{equation} \label{eq:derivativepsi2}
\begin{split}
\frac{\partial}{\partial u} \left( \frac{\psi(u,\rho)}{u} \right) &= \frac{1}{u^2} \left( -\ln(1-u) -\rho \ln(\rho) \right. \\
&\;\;\;\;- \left. (1-\rho) \ln(1-\rho) \right. \\
&\;\;\;\;+ \left. (1-\rho) \ln (1-\rho-\tilde{\gamma}_u ) \right. \\
&\;\;\;\;+ \left. \rho \ln(\rho+\tilde{\gamma}_u-u) \right) \\
&= \frac{1}{u^2} \left( -\ln(1-u)+\rho \ln \left( \frac{\rho+\tilde{\gamma}_u-u}{\rho} \right) \right. \\
&\;\;\;\;+ \left. (1-\rho) \ln \left( \frac{1-\rho-\tilde{\gamma}_u}{1-\rho} \right) \right).
\end{split}
\end{equation}
The function $-\ln(x)$ is convex, and Jensen's inequality gives
\begin{equation} \label{eq:jensen}
\begin{split}
-\ln(1-u) &= -\ln \left( \rho \frac{\rho+\tilde{\gamma}_u-u}{\rho} + (1-\rho) \frac{1-\rho-\tilde{\gamma}_u}{1-\rho} \right) \\
&\leq  - \rho \ln \left( \frac{\rho+\tilde{\gamma}_u-u}{\rho} \right) \\
&\;\;\;\;- (1-\rho) \ln \left( \frac{1-\rho-\tilde{\gamma}_u}{1-\rho} \right)
\end{split}
\end{equation}
from which it follows (by substituting the upper bound from (\ref{eq:jensen}) into (\ref{eq:derivativepsi2})) that $\frac{\partial}{\partial u} \left( \frac{\psi(u,\rho)}{u} \right) \leq 0$, and the result follows.

\subsection{Proof of Property 4} \label{sec:property4}

Calculating the partial derivative of the objective function in (\ref{eq:f})  with respect to $\gamma$ and setting it equal to zero results in the equation in (\ref{eq:derivativegamma1}), which can be simplified to
\begin{equation} \label{eq:zero}
\begin{split}
&-q^2 \gamma^3 + (-4+u q^2-\rho q^2 +4q)\gamma^2 \\
&+ (-4uq-u^2q+u^2+4u-4\rho u+4 \rho uq)\gamma \\
&-u^2-\rho u^2q+\rho u^2+u^2q = 0.
\end{split}
\end{equation}
Setting $u=0$ in (\ref{eq:zero}), gives
\begin{equation} \notag 
\begin{split}
&-q^2 \gamma^3 + (-4-\rho q^2 +4q)\gamma^2  = 0
\end{split}
\end{equation}
with solutions
\begin{displaymath}
\gamma = \begin{cases}
0, \\
0, \\
\frac{-4-\rho q^2+4q}{q^2}. \end{cases}
\end{displaymath}
Since $\tilde{\gamma}_u$ is upper-bounded by $u/2$ and lower-bounded by $0$ (see (\ref{eq:f})), it follows that $\tilde{\gamma}_0=\tilde{\gamma}(u=0)=0$. Now, the limit
\begin{equation} \label{eq:limit}
\begin{split}
\lim_{u \longrightarrow 0} \frac{\psi(u,\rho)}{u} &= \lim_{u \longrightarrow 0} \frac{\partial \psi(u,\rho)}{\partial u} \\
&= \ln \left( \frac{q-2}{q-1} \right)
+\lim_{u \longrightarrow 0} \ln \left( \frac{\left( \rho+\tilde{\gamma}_u-u \right) u}{\left( u -2 \tilde{\gamma}_u \right) (1-u)} \right) \\
&= \ln \left( \frac{q-2}{q-1} \right) \\
&\;\;\;\;+\lim_{u \longrightarrow 0} \ln \left( \frac{ \rho-2u+\tilde{\gamma}_u+\tilde{\gamma}_u' u }{1-2\tilde{\gamma}_u'-2u +2\tilde{\gamma}_u +2\tilde{\gamma}_u' u} \right) \\
&= \ln \left( \frac{q-2}{q-1} \right) + \ln \left( \frac{\rho}{1-2\tilde{\gamma}_0' } \right) \\
&= \ln \left( \frac{(q-2)\rho}{(q-1)(1-2\tilde{\gamma}_0') } \right) \\
\end{split}
\end{equation}
where the first and third equalities follow from l'H\^opital's rule, the second equality follows from (\ref{eq:12}), and the fourth equality follows under the assumption that $1-2 \tilde{\gamma}_0'$ is nonzero. 
We will now show that this is indeed the case.

Since $\tilde{\gamma}_0=0$, we may write $\tilde{\gamma}_u = \tilde{\gamma}'_0 u +O(u^2)$ (Taylor series expansion around $u=0$). Substituting  $\tilde{\gamma}'_0 u +O(u^2)$ for $\gamma$ in (\ref{eq:zero}) and taking the limit as $u$ approaches zero, we get
\begin{equation} \label{eq:3}
\begin{split}
\left(-4-\rho q^2
+4q \right) (\tilde{\gamma}_0')^2+4 \left(q-1 \right) \left(\rho-1 \right)  \tilde{\gamma}_0' &\\
-1-\rho q + \rho +q &= 0
\end{split}
\end{equation}
which has the solutions (when $-4-\rho q^2+4q$ is assumed to be nonzero)
\begin{equation} \notag 
\tilde{\gamma}_0' = \frac{-2(1-\rho)(q-1) \pm (q-2)\sqrt{\rho(1-\rho)(q-1)}}{4+\rho q^2-4 q}
\end{equation}
from which it follows that
\begin{equation} \label{eq:11}
1-2\tilde{\gamma}_0' = \frac{\rho(q-2)^2 \pm 2(q-2)\sqrt{\rho(1-\rho)(q-1)}}{4+\rho q^2-4 q}.
\end{equation}
Now, inserting this expression into (\ref{eq:limit}), we get
\begin{equation} \label{eq:limit1}
\begin{split}
&\lim_{u \longrightarrow 0} \frac{\psi(u,\rho)}{u} \\
&= \ln \left( \frac{\rho( 4+\rho q^2-4 q)}{(q-1) \left( \rho(q-2) \pm 2\sqrt{\rho(1-\rho)(q-1)} \right)} \right).
\end{split}
\end{equation}
Setting
\begin{displaymath}
\frac{\rho( 4+\rho q^2-4 q)}{(q-1) \left( \rho(q-2) \pm 2\sqrt{\rho(1-\rho)(q-1)} \right)} = 1
\end{displaymath}
results in the equation
\begin{equation} \label{eq:1}
\rho \left( -(q-1)(q+2) + \rho q^2 \right) = \pm 2 (q-1) \sqrt{\rho (1-\rho) (q-1)}.
\end{equation}
Squaring both sides of the equality in (\ref{eq:1}), we get (after some re-arrangement)
\begin{equation} \label{eq:2}
\begin{split}
&q^4 \rho^3-2 q^2(q+2)(q-1) \rho^2 \\
&+\left( (q-1)^2(q+2)^2+4(q-1)^3 \right) \rho
- 4(q-1)^3=0
\end{split}
\end{equation}
with solutions
\begin{displaymath}
\rho = \begin{cases}
(q-1)/q, \\
(q-1)/q,  \\
4(q-1)/q^2.
\end{cases}
\end{displaymath}
The third solution is not valid, since we have assumed $-4-\rho
q^2+4q$ to be nonzero. Thus, (\ref{eq:2}) has only a single
solution. To prove that the limit in (\ref{eq:limit1}) is strictly
less than zero for $\rho < (q-1)/q$ when $-4-\rho q^2+4q$ is
nonzero, it is sufficient to evaluate the limit for some $\rho<
(q-1)/q$ such that $-4-\rho q^2+4q$ is nonzero. In this respect, we
choose $\rho = (q-1)/q^2$, which is strictly smaller than $\min
\left( (q-1)/q, 4(q-1)/q^2 \right)$. Since the denominator in
(\ref{eq:11}) is negative for this value of $\rho$, it follows from
(\ref{eq:11}) (the $\pm$ will be a minus) that
\begin{displaymath}
\begin{split}
\frac{(q-2)\rho}{(q-1)(1-2\tilde{\gamma}_0')} &= \frac{\rho(4+\rho q^2-4 q)}{(q-1)(\rho(q-2) - 2\sqrt{\rho(1-\rho)(q-1)})} \\
&= \frac{3}{2 \sqrt{q^2-q+1}-(q-2)} \\
&< \frac{3}{2 \sqrt{(q-1)^2}-(q-2)} = \frac{3}{q}
\end{split}
\end{displaymath}
where the strict inequality follows from the fact that $q^2-q+1 > q^2-2q+1 = (q-1)^2$. Now, since $q \geq 3$,
\begin{displaymath}
\lim_{u \longrightarrow 0} \frac{\psi(u,\rho)}{u} = \ln \left( \frac{(q-2)\rho}{(q-1)(1-2\tilde{\gamma}_0')} \right) < 0
\end{displaymath}
for $\rho = (q-1)/q^2$.

Finally, when  $-4-\rho q^2+4q$ is zero, (\ref{eq:3}) reduces to
\begin{equation} \notag 
4 \left(q-1 \right) \left(\rho-1 \right)  \tilde{\gamma}_0'-1-\rho q + \rho +q = 0
\end{equation}
with solution $1-2\tilde{\gamma}_0' = 1/2$, from which it follows from (\ref{eq:limit}) that
\begin{displaymath}
\lim_{u \longrightarrow 0} \frac{\psi(u,\rho)}{u} = \ln \left( \frac{8 (q-2)}{q^2} \right) < 0
\end{displaymath}
for $q \geq 5$. Note that for $q=3$ and $4$, $4(q-1)/q^2 \geq (q-1)/q$, and we can conclude that Property 4 is proved.




\balance



\end{document}